\documentclass[aps,prd,twocolumn,superscriptaddress,preprintnumbers]{revtex4-2}

\usepackage{graphicx}
\usepackage{dcolumn}
\usepackage{bm}
\usepackage{epstopdf}
\usepackage[caption=false]{subfig}

\usepackage{amsmath} 
\newcommand{\angstrom}{\textup{\AA}}

\newcommand{\cevns}{\protect{CE$\nu$NS}\,}

\newcommand{\iso}[2]{\protect{\ensuremath{{}^{#1}\textrm{#2}}}\,}

\usepackage{verbatim}

\begin{document}
\preprint{LA-UR-21-24983}

\title{First Dark Matter Search Results From Coherent CAPTAIN-Mills} 

\affiliation{Bartoszek~Engineering,~Aurora,~IL~60506,~USA}
\affiliation{Columbia~University,~New~York,~NY~10027,~USA}
\affiliation{University~of~Edinburgh,~Edinburgh,~United~Kingdom}
\affiliation{Embry$-$Riddle~Aeronautical~University,~Prescott,~AZ~86301,~USA }
\affiliation{University~of~Florida,~Gainesville,~FL~32611,~USA}
\affiliation{Los~Alamos~National~Laboratory,~Los~Alamos,~NM~87545,~USA}
\affiliation{Massachusetts~Institute~of~Technology,~Cambridge,~MA~02139,~USA}
\affiliation{Universidad~Nacional~Aut\'{o}noma~de~M\'{e}xico,~CDMX~04510,~M\'{e}xico}
\affiliation{University~of~New~Mexico,~Albuquerque,~NM~87131,~USA}
\affiliation{New~Mexico~State~University,~Las~Cruces,~NM~88003,~USA}
\affiliation{Texas~A$\&$M~University,~College~Station,~TX~77843,~USA}

\author{A.A.~Aguilar-Arevalo}
\affiliation{Universidad~Nacional~Aut\'{o}noma~de~M\'{e}xico,~CDMX~04510,~M\'{e}xico}
\author{D.~S.\,M.~Alves}
\affiliation{Los~Alamos~National~Laboratory,~Los~Alamos,~NM~87545,~USA}
\author{S.~Biedron}
\affiliation{University~of~New~Mexico,~Albuquerque,~NM~87131,~USA}
\author{J.~Boissevain}
\affiliation{Bartoszek~Engineering,~Aurora,~IL~60506,~USA}
\author{M.~Borrego}
\affiliation{Los~Alamos~National~Laboratory,~Los~Alamos,~NM~87545,~USA}
\author{M.~Chavez$-$Estrada}
\affiliation{Universidad~Nacional~Aut\'{o}noma~de~M\'{e}xico,~CDMX~04510,~M\'{e}xico}
\author{A.~Chavez}
\affiliation{Los~Alamos~National~Laboratory,~Los~Alamos,~NM~87545,~USA}
\author{J.M.~Conrad}
\affiliation{Massachusetts~Institute~of~Technology,~Cambridge,~MA~02139,~USA}
\author{R.L.~Cooper}
\affiliation{Los~Alamos~National~Laboratory,~Los~Alamos,~NM~87545,~USA}
\affiliation{New~Mexico~State~University,~Las~Cruces,~NM~88003,~USA}
\author{A.~Diaz}
\affiliation{Massachusetts~Institute~of~Technology,~Cambridge,~MA~02139,~USA}
\author{J.R.~Distel}
\affiliation{Los~Alamos~National~Laboratory,~Los~Alamos,~NM~87545,~USA}
\author{J.C.~D'Olivo}
\affiliation{Universidad~Nacional~Aut\'{o}noma~de~M\'{e}xico,~CDMX~04510,~M\'{e}xico}
\author{E.~Dunton}
\affiliation{Columbia~University,~New~York,~NY~10027,~USA}
\author{B.~Dutta}
\affiliation{Texas~A$\&$M~University,~College~Station,~TX~77843,~USA}
\author{A.~Elliott}
\affiliation{Embry$-$Riddle~Aeronautical~University,~Prescott,~AZ~86301,~USA }
\author{D.~Evans}
\affiliation{Los~Alamos~National~Laboratory,~Los~Alamos,~NM~87545,~USA}
\author{D.~Fields}
\affiliation{University~of~New~Mexico,~Albuquerque,~NM~87131,~USA}
\author{J.~Greenwood}
\affiliation{Embry$-$Riddle~Aeronautical~University,~Prescott,~AZ~86301,~USA }
\author{M.~Gold}
\affiliation{University~of~New~Mexico,~Albuquerque,~NM~87131,~USA}
\author{J.~Gordon}
\affiliation{Embry$-$Riddle~Aeronautical~University,~Prescott,~AZ~86301,~USA }
\author{E.~Guarincerri}
\affiliation{Los~Alamos~National~Laboratory,~Los~Alamos,~NM~87545,~USA}
\author{E.C.~Huang}
\affiliation{Los~Alamos~National~Laboratory,~Los~Alamos,~NM~87545,~USA}
\author{N.~Kamp}
\affiliation{Massachusetts~Institute~of~Technology,~Cambridge,~MA~02139,~USA}
\author{C.~Kelsey}
\affiliation{Los~Alamos~National~Laboratory,~Los~Alamos,~NM~87545,~USA}
\author{K.~Knickerbocker}
\affiliation{Los~Alamos~National~Laboratory,~Los~Alamos,~NM~87545,~USA}
\author{R.~Lake}
\affiliation{Embry$-$Riddle~Aeronautical~University,~Prescott,~AZ~86301,~USA }
\author{W.C.~Louis}
\affiliation{Los~Alamos~National~Laboratory,~Los~Alamos,~NM~87545,~USA}
\author{R.~Mahapatra}
\affiliation{Texas~A$\&$M~University,~College~Station,~TX~77843,~USA}
\author{S.~Maludze}
\affiliation{Texas~A$\&$M~University,~College~Station,~TX~77843,~USA}
\author{J.~Mirabal}
\affiliation{Los~Alamos~National~Laboratory,~Los~Alamos,~NM~87545,~USA}

\author{R.~Moreno}
\affiliation{Embry$-$Riddle~Aeronautical~University,~Prescott,~AZ~86301,~USA }
\author{H.~Neog}
\affiliation{Texas~A$\&$M~University,~College~Station,~TX~77843,~USA}
\author{P.~deNiverville}
\affiliation{Los~Alamos~National~Laboratory,~Los~Alamos,~NM~87545,~USA}
\author{V.~Pandey}
\affiliation{University~of~Florida,~Gainesville,~FL~32611,~USA}
\author{J.~Plata$-$Salas}
\affiliation{Universidad~Nacional~Aut\'{o}noma~de~M\'{e}xico,~CDMX~04510,~M\'{e}xico}
\author{D.~Poulson}
\affiliation{Los~Alamos~National~Laboratory,~Los~Alamos,~NM~87545,~USA}
\author{H.~Ray}
\affiliation{University~of~Florida,~Gainesville,~FL~32611,~USA}
\author{E.~Renner}
\affiliation{Los~Alamos~National~Laboratory,~Los~Alamos,~NM~87545,~USA}
\author{T.J.~Schaub}
\affiliation{University~of~New~Mexico,~Albuquerque,~NM~87131,~USA}
\author{M.H.~Shaevitz}
\affiliation{Columbia~University,~New~York,~NY~10027,~USA}
\author{D.~Smith}
\affiliation{Embry$-$Riddle~Aeronautical~University,~Prescott,~AZ~86301,~USA }
\author{W.~Sondheim}
\affiliation{Los~Alamos~National~Laboratory,~Los~Alamos,~NM~87545,~USA}
\author{A.M.~Szelc}
\affiliation{University~of~Edinburgh,~Edinburgh,~United~Kingdom}
\author{C.~Taylor}
\affiliation{Los~Alamos~National~Laboratory,~Los~Alamos,~NM~87545,~USA}
\author{W.H.~Thompson}
\affiliation{Los~Alamos~National~Laboratory,~Los~Alamos,~NM~87545,~USA}
\author{M.~Tripathi}
\affiliation{University~of~Florida,~Gainesville,~FL~32611,~USA}
\author{R.T.~Thornton}
\affiliation{Los~Alamos~National~Laboratory,~Los~Alamos,~NM~87545,~USA}
\author{R.~Van~Berg}
\affiliation{Bartoszek~Engineering,~Aurora,~IL~60506,~USA}
\author{R.G.~Van~de~Water}
\affiliation{Los~Alamos~National~Laboratory,~Los~Alamos,~NM~87545,~USA}
\author{S.~Verma}
\affiliation{Texas~A$\&$M~University,~College~Station,~TX~77843,~USA}
\author{K.~Walker}
\affiliation{Embry$-$Riddle~Aeronautical~University,~Prescott,~AZ~86301,~USA }

\collaboration{The CCM Collaboration}

\begin{abstract}
This paper describes the operation of the Coherent CAPTAIN-Mills (CCM) detector located at the Los Alamos Neutron Science Center (LANSCE) at Los Alamos National Laboratory (LANL).  CCM is a 10-ton liquid argon (LAr) detector located 20 meters from a high flux neutron/neutrino source and is designed to search for sterile neutrinos ($\nu_s$'s) and light dark matter (LDM).  An engineering run was performed in Fall 2019 to study the characteristics of the CCM120 detector by searching for coherent scattering signals consistent with $\nu_s$'s and LDM resulting from the production and decays of $\pi^+$ and $\pi^0$ in the tungsten target. New parameter space in a leptophobic dark matter (DM) model was excluded for DM masses between $\sim2.0$ and 30\,MeV. The lessons learned from this run have guided the development and construction of the new CCM200 detector that will begin operations in 2021 and significantly improve on these searches.
\end{abstract}

\maketitle




\section{Introduction} 

The LANSCE facilty provides a unique opportunity to search for LDM and confirm whether $\nu_s$'s, exist and mix with the three active neutrinos ($\nu_e$, $\nu_{\mu}$, $\nu_{\tau}$) in the Standard Model (SM). The composition of DM\,\cite{Tanabashi:2018dm} is one of the most important issues in physics today, although the mass and properties of DM are presently unknown. The Lujan facilty located inside LANSCE is home to a high intensity pulsed neutron/neutrino source. LDM can be produced from $\pi^0$ decay in the Lujan beam dump and detected by scattering in the CCM detector. In addition, by measuring neutrino scattering in the cryostat, a search can be made for $\nu_s$'s\,\cite{Sterilenu} which, if they exist, would have a profound impact on our understanding of particle physics and deep implications for cosmology.  The discovery of $\nu_s$'s would be a concrete realization of physics Beyond the Standard Model (BSM) and indicate possible connections to the dark sector (i.e., the dynamical sector associated with DM in the Universe\,\cite{Tanabashi:2018sn}).  

The CCM experiment makes use of the intense pion production at the Lujan facility to search for DM from $\pi^0$ decay and to measure mono-energetic 30\,MeV muon-neutrinos ($\nu_{\mu}'s$) from $\pi^+$ decay\,\cite{vdw:2019aps}. CCM is designed to measure Coherent Elastic Neutrino-Nucleus Scattering (\cevns) with an instrumented 10-ton liquid argon scintillation detector. In the Lujan beam dump, LDM can be produced by $\pi^0$ decay.  The CCM detector will measure LDM with a higher \cevns rate than expected.  At the same time CCM will measure a rate lower than expected if $\nu_{\mu}$ oscillates into $\nu_s$. Since LDM from $\pi^0$ decay will be more energetic than $\nu_s$'s from neutrino oscillations, the two effects can be separated, so that the CCM detector will be able to search for both LDM and sterile neutrinos simultaneously. However, in this paper our analysis is focused on searching for LDM events having a coherent nuclear scattering signature with energies $>50$\,keV. 

In the remainder of this section we introduce the physics goals for the CCM detector (i.e., the search for $\nu_s$'s and LDM).  In Sec.~\ref{sec:ccm120detector}, we describe the design considerations for constructing the CCM120 detector containing 120 8-inch photomultiplier tubes (PMTs).  In Sec.~\ref{sec:particlesources}, the beam, target and particle sources for neutrinos and LDM are presented.  In Sec.~\ref{sec:dataanalysis}, we introduce new analysis techniques (e.g., event building techniques and the Optical Model) and apply them to the data from the engineering run.  These techniques are used to develop data selection criteria for the LDM search.  And finally, in Sec.~\ref{sec:cl} we present the LDM mass region excluded in the leptophobic model by applying the new analysis techniques to the CCM120 data.  A note to the reader, we use natural units throughout this paper $(\hbar=c=1)$.

\subsection{Physics Goals for the CCM detector}
The two main physics goals of CCM are to search for LDM coming from $\pi^0$ decay in flight in the beam dump and to search for neutral current disappearance (the $\nu_s$ signature) of mono-energetic $\nu_{\mu}$ coming from $\pi^+$ decay at rest in the beam dump.  For the DM search, CCM will look for an excess of coherent scattering events with recoil Ar nuclei energy greater than $\sim$50\,keV while for the $\nu_s$ search, CMM will look for a deficit of $\nu_{\mu}$ \cevns events with recoil Ar nuclei energy less than $\sim$50\,keV (the maximum recoil energy of the Ar nucleus for a 30 MeV neutrino). 
In both cases the coherent scattering events will be prompt and will coincide approximately with the time of the beam spill. Both LDM and $\nu_{\mu}$ are assumed to travel at nearly the speed of light, and their coherent interactions in CCM will generally occur before the arrival of the beam neutron background.  Note that none of the $\nu_s$'s, but only some of the DM particles, will produce recoil Ar nuclei energies with less than 50 keV.

\subsubsection{Coherent Elastic Neutrino-Nucleus Scattering} 
\cevns~was first suggested as a probe for the weak current in 1974\,\cite{Freedman:1973yd} soon after the experimental discovery of weak neutral current in neutrino interactions. \cevns~is flavor blind at the tree level, meaning that each SM neutrino \cite{Tanabashi:2018sm} interacts identically, and its scattering rate is a few orders of magnitude larger than other competing neutrino scattering events at energies \begin{math}\mathcal{O}\end{math}(10 MeV). Despite the rate enhancement, observing \cevns~is difficult due to experimental challenges such as low nuclear recoil energies \begin{math}\mathcal{O}\end{math}(10 keV), and being the only experimental signature, it requires intense sources and large target masses.  \cevns eluded detection for over four decades until it was first observed in 2017 in CsI\,\cite{Akimov:2017ade}, and more recently in Ar\,\cite{Akimov:2020pdx} by the \mbox{COHERENT} collaboration.

In \cevns, the neutrino interacts with the whole nucleus and the collection of individual nucleons behaves coherently. The scattering results in a relatively large cross section with the nucleus remaining in its ground state. The experimental signature for \cevns is low-energy nuclear recoils $T$ of $\mathcal{O}$(10 keV), and the differential cross section for the process in terms of recoil energy is written as:
\begin{equation}\label{Eq:xs_recoil}
\frac{\mathrm{d}\sigma}{ \mathrm{d}T} = \frac{G^{2}_{F}}{\pi} M_{A} \left(1-\frac{T}{E_{\nu}}-\frac{M_A T}{2 E^2_\nu}\right)~\frac{Q^2_{W}}{4}~F_{W}^2(Q^2), 
\end{equation}
where $G_F$ is the Fermi coupling constant, $E_\nu$ is the energy of the incoming neutrino and $M_A$ is the nuclear target mass. The recoil energy $T=E_R=q^2/(2M_{A})$ has values in the range $[0,2E_\nu^2/(M_{A}+2E_\nu)]$. $Q_W$ is the weak nuclear charge given as $Q_{W} = [g_p^V Z+g_n^V N] = [(1-4\sin^2\theta_{\text{W}})Z-N]$ where $N(Z)$ are neutron (proton) numbers and $\theta_{\mathrm{W}}$ is the weak mixing angle. Since the weak-interaction charge of the proton is suppressed due to the small value of $(1-4\sin^2\theta_{\text{W}})$, the \cevns rate primarily depends upon the square of the number of neutrons. 

$F_{\text{w}}(q^2)$ is the weak form factor of the target nucleus. To first approximation, $F_{\text{w}}(q^2)$ depends on the nuclear density distribution of protons and neutrons. In the coherence limit $q^2\to 0$, it is normalized to $F_{\text{w}}(0)=1$. The coherent enhancement of the cross section is reflected by the $N^2$ scaling via the weak charge, given the suppression of the proton weak charge. 

The CCM experiment with its proposed energy threshold of 20\,keVnr (nuclear recoil) combined with Lujan Center’s intense neutrino source presents a powerful avenue to detect coherent scattering off the argon nucleus. Understanding \cevns on argon is also important for both DM and sterile neutrino searches since coherent elastic scattering is the detection signal for both searches, and for neutrinos is flavor independent.

\subsubsection{Current Status of SBN Oscillations and the Sterile Neutrino Hypothesis}
Over the last three decades, a series of solar, atmospheric, reactor, and accelerator neutrino oscillation experiments have proven the existence of neutrino oscillations among the three active neutrinos, implying that neutrinos have mass and that SM must be extended \cite{Nu_mass}.  
The LSND and MiniBooNE experiments have published strong evidence (6.1\,$\sigma$ significance) for muon-neutrino to electron-neutrino appearance oscillations at $\Delta m^2\sim1$\,eV$^2$\,\cite{Aguilar-Arevalo:2020nvw}. In addition, there is evidence for electron-neutrino disappearance into $\nu_s$'s from radioactive sources and reactor neutrino experiments.  There is also weak evidence from the IceCube experiment for muon-neutrino disappearance into $\nu_s$'s at the same $\sim 1$ eV$^2$ mass scale 
\cite{BEST,Aartsen:2020iky}. However, other experiments, such as KARMEN and MINOS+\,\cite{KARMEN,Adamson:2020jvo}, have not observed evidence for $\nu_s$'s. In order to prove whether $\nu_s$'s exist, it is crucial to (i)\,improve the sensitivity of muon-neutrino oscillation experiments and (ii)\,observe the oscillations as a function of $L/E$, 
$P=\sin^2(2\theta) \sin^2(1.27 \Delta m^2 L/E)$,
where $P$ is the oscillation probability, $\theta$ is the mixing angle, $\Delta m^2$ is the difference in neutrino eigenmasses squared~(eV$^{\mathrm{2}}$), $L$ is the distance traveled by the neutrino~(m) and $E$ is the neutrino energy~(MeV). The CCM experiment has the capability of doing both by using the 30-MeV mono-energetic muon neutrinos from stopped $\pi^+$ decay in the Lujan beam-dump and by measuring $\nu_{\mu}$ scattering coherently off argon nuclei in the CCM LAr cryostat. With a neutrino energy of 30 MeV and a mass scale of $\sim$1~eV$^2$, muon-neutrino disappearance oscillations would be expected over distances of tens of meters. In addition, neutrino-nucleus coherent scattering is a neutral-current process, where active neutrinos have the same cross sections while $\nu_s$'s do not interact at all. Therefore, the observation of neutrino oscillations via a neutral-current reaction, but with a reduced neutrino flux, would prove the existence of $\nu_s$'s.

\subsubsection{Light Dark Matter and Current Status}

From the Planck data \cite{Planck:2020}, approximately 85$\%$ of the matter density of the universe is made up of electromagnetically undetectable, gravitationally interacting DM. 
Despite decades of effort, only indirect evidence for DM has been observed.  Meanwhile, direct experimental measurements of DM remain elusive~\cite{Tanabashi:2018dms}.   
Current searches for thermal relic DM 
have low sensitivity below $\sim$1~GeV mass due to detector thresholds; however, thermal DM below this mass does not sufficiently annihilate in the early universe to reproduce the measured relic abundance. The annihilation rate $Y$ depends on the DM mass $m_{_\mathrm{DM}}$ as shown in the following equation
\begin{equation}
    Y=\frac{g_{_\mathrm{D}}^2 g_{_{\mathrm{SM}}}^2}{16 \pi^2} \left(\frac{m_{_\mathrm{DM}}}{m_{_\mathrm{MED}}}\right)^4 
     =\varepsilon^2\alpha_{_\mathrm{D}}\left(m_{\chi}/m_{_\mathrm{V}}\right)^4,
\end{equation}
where $m_{_{\mathrm{MED}}}$ is the mass of the mediator, $g_{_\mathrm{D}}$ is the strength of the coupling between the mediator and the DM candidate, $g_{_{\mathrm{SM}}}$ is the strength of the coupling between the mediator and the SM states into which the DM is annihilating \cite{Izaguirre:2015yja}. If the mediator is the $Z$-boson, this suggests that a DM mass between $\sim$1 and 1000~GeV is required to generate the observed DM relic density, and is the origin of the Lee-Weinberg bound \cite{Lee:1977ua}. DM models with interactions mediated by $Z$ exchange have become increasingly restricted by WIMP searches in this mass range \cite{Escudero:2016gzx}, thus adding to the motivation for LDM searches.
The Lee-Weinberg bound can be circumvented with the introduction of new light-mass mediators - reducing the relevant $m_{_\mathrm{MED}}$~\cite{Boehm:2003hm,Fayet:2004bw,Pospelov:2007mp,Holdom:1985ag,deNiverville:2016rqh}. There are many sub-GeV DM models (e.g. \cite{Fayet:1990wx,Boehm:2003hm,Fayet:2004bw,Pospelov:2007mp,Batell:2014yra,Izaguirre:2017bqb,Jordan:2018gcd,Magill:2018tbb}. See \cite{Alexander:2016aln,Battaglieri:2017aum} for recent community reviews on the topic), but the vector portal model is the simplest model and will be used to characterize sub-GeV DM in the following search. The vector portal model allows for a $U(1)’$ gauge boson $V$ $(m_{_V}=m_{_{\mathrm{MED}}})$ which can be described as a `dark photon' to kinetically mix with the SM photon with a mixing strength of $\varepsilon = g_{_\mathrm{SM}}$. The assumed complex scalar DM $\chi$ ($m_{\chi} = m_{_\mathrm{DM}}$) couples to the vector mediator with dark gauge coupling $g_{_\mathrm{D}}$ with $\alpha_{_\mathrm{D}} = g_{_\mathrm{D}}^2/4\pi$.

\begin{figure}
    \centering
    \includegraphics[width=0.48\textwidth]{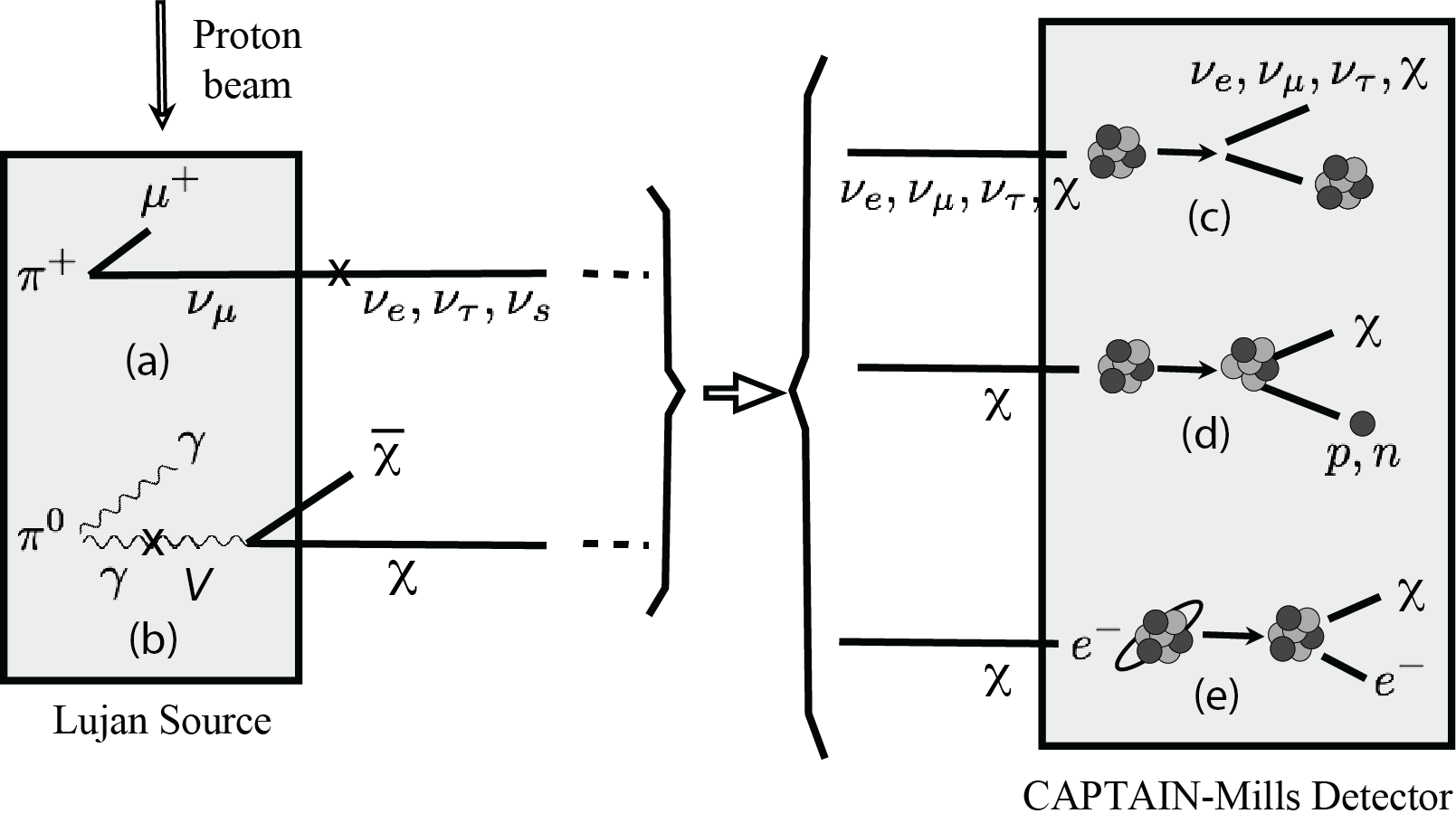}
    \caption{An 800-MeV proton beam strikes a tungsten target that produces (a)\,charged pions that decay into muon neutrinos, and (b)\,neutral pions that may decay to sub-GeV dark matter $\chi$.  The neutrinos and dark matter particles travel to the CCM detector where they are detected by (c)\,\cevns, (d)\,quasi-elastic nucleon scattering, or (e)\,elastic electron scattering.  Active neutrinos from $\nu_{\mu}$ oscillations produce nuclear recoils ($<$50\,keV), while $\nu_s$'s do not.  Likewise, the LDM particles ($\chi$'s) will produce energetic nuclear recoils ($>$50\,keV).}
    \label{fig:LDMneutrinoScatter}
\end{figure}

The CCM experiment searches for DM across a wide spectrum of sub-GeV masses and other potential parameters using the vector portal model. The Lujan source would produce vector portal DM through similar channels as neutrinos, so the experiment has the advantage of timing for background rejection. On the other end, this DM would interact through the same coherent scattering channel as \cevns (see Fig.~\ref{fig:LDMneutrinoScatter}). Thus CCM would not need to alter its approach, merely look for recoil energies above the neutrino limit. After energy and timing selection criteria are applied, only random backgrounds would remain; this includes backgrounds out-of-time with the beam (e.g., cosmic, thermal neutron bath, radioactive material, etc.). It should be noted that CCM is situated in a neutron research facility with experiments nearby using the neutron beams.  There are 5\,m of steel and 1\,m of concrete immediately surrounding the tungsten target with additional concrete and steel between the source and the CCM detector.  This reduces the flux and energy of neutrons near the CCM detector.  Furthermore, the CCM detector is located 90$^{\mathrm{o}}$ with respect to the incident proton beam.
\\ 

\section{Description of the CCM120 Detector \label{sec:ccm120detector}}
The CCM detector was designed to directly probe the LSND result by measuring $\nu_{\mu}$ disappearance as a function of distance, and to search for low mass DM through coherent scattering, as well as other potential dark sector particles.

\subsection{Overall Design Considerations and Goals}

\begin{figure}
    \centering
    \includegraphics[width=0.48\textwidth]{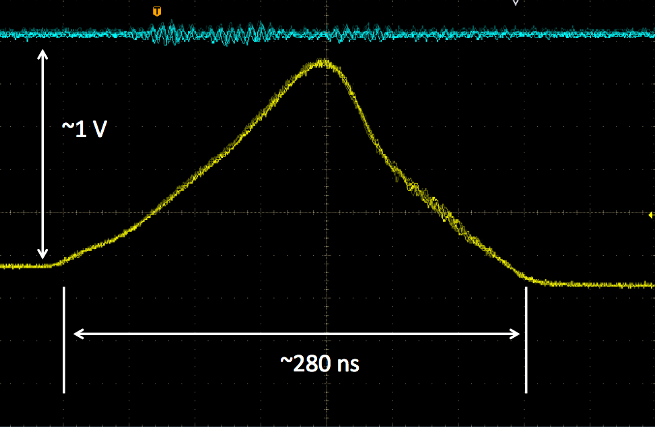}
    \caption{The time profile of an individual proton bunch incident on the Lujan tungsten production target as measured by a pick-up coil placed just upstream of the target.}
    \label{fig:ProtonBunchTime}
\end{figure}

The LANSCE Lujan target provides an estimated 4$\pi$ flux of 4.74$\times$10$^5$ $\nu$/cm$^2$/s for each neutrino species ($\nu_\mu$ from $\pi^+$ decay; $\bar{\nu}_{\mu}$ and $\nu_e$ from the decay of the daughter $\mu^+$) at 20\,m from the production target for nominal beam conditions.
 The neutrinos are produced from $\pi^+$ decays when an 800\,MeV bunched proton beam from the proton storage ring (PSR) strikes the tungsten target with a repetition rate of 20\,Hz.
 The time profile of an individual proton bunch at the Lujan center is shown in Fig.~\ref{fig:ProtonBunchTime}.  Each bunch consists of $\sim 3.1\,\times\,10^{13}$ protons with a triangular time distribution spanning $\sim280$\,ns. 

\begin{figure}
    \centering
    \includegraphics[width=0.50\textwidth]{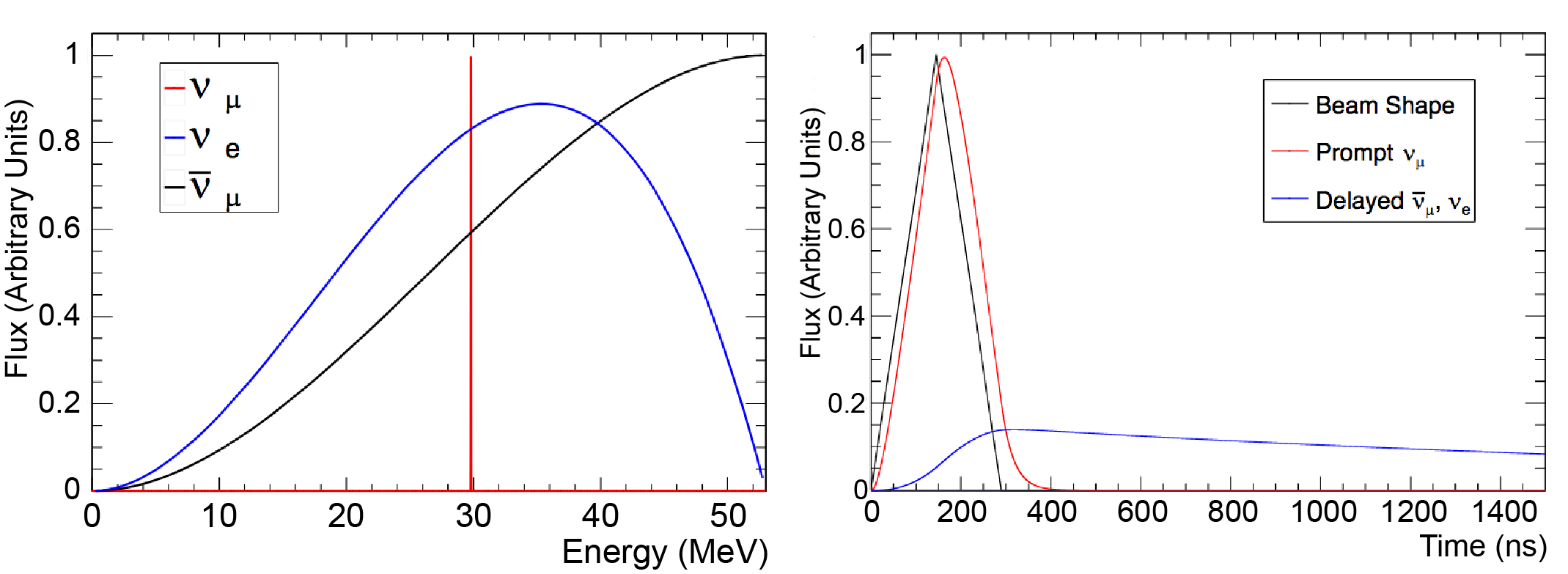}
    \caption{Energy spectra for individual neutrino species (left) resulting from $\pi^+$ and $\mu^+$ decays, and simulated time profiles (right) of the neutrinos from an ideal stopped-pion neutrino source.}
    \label{fig:TimeEnergyNeutrinos}
\end{figure}

\begin{figure}
    \centering
    \includegraphics[width=0.48\textwidth]{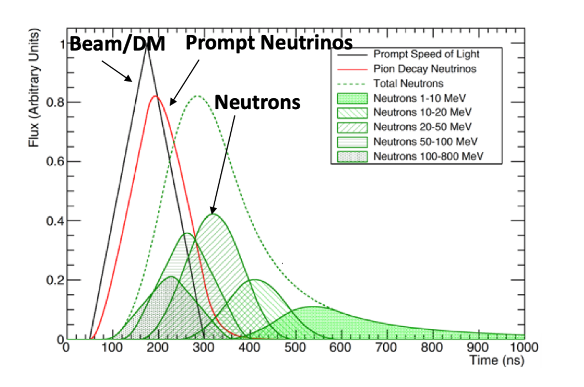}
    \caption{The relative prompt neutrino and DM timing spectra are shown and compared to an exponential energy spectrum of neutrons for the case where the detector is placed 20\,m from the Lujan target with a 280\,ns wide (at the base) beam.  Time of flight measurements outside CCM120 determined the fastest neutrons were about 20\,MeV. A beam time profile of 100\,ns, a planned near term PSR upgrade, would yield a region that is only populated by mono-energetic 30\,MeV stopped pion neutrinos and potential DM. }
    \label{fig:NeutronTime}
\end{figure}

The energy spectrum and time distribution of these beam neutrinos are shown in \mbox{Fig.\,\ref{fig:TimeEnergyNeutrinos}}.
Neutrons are also produced at the Lujan target and constitute the primary source of background for CCM. Thanks to their high mass and the moderating material around the production target, neutrons reach CCM later than neutrinos and DM, providing a $>$100\,ns background free window that enables the search proposed here. The simulated arrival times of the beam neutrinos, DM, and beam neutrons at CCM are shown in Fig.\,\ref{fig:NeutronTime} assuming a baseline $L$=20 m.  Time of flight measurements using multiple \mbox{Eljen} EJ301 detectors determined that the fastest neutrons outside the CCM120 detector were about 20\,MeV. 

In order to search for 30-MeV muon-neutrino disappearance into $\nu_s$'s at Lujan, the detector must be able to measure low-energy nuclear recoils from neutrino-nucleus coherent scattering. A LAr cryostat with PMT light detection is ideal for this purpose, as it is fairly inexpensive, has low radioactivity, and produces a high yield of scintillation light.
The LAr cryostat should also be movable for mapping out the oscillation pattern over distances of tens of meters.
For 30-MeV muon-neutrinos interacting on Ar nuclei, the maximum recoil energy is about 50 keV.
We plan to achieve an energy threshold of 10\,keV and an overall light production/collection efficiency of 1 photoelectron/keV to collect on average 10 photoelectrons at the threshold.
 This provides an energy resolution $<20\%$ at the 50\,keV endpoint. Also, the detector mass  needs to be $>3$ tons to be able to observe more than 5 \cevns events per day, but it cannot be excessively large to preserve the capability of being moved and operated at different baselines.
 Additionally, the neutron background in the beam-spill time window will need to be less than 5 events per day, so that the signal-to-background is greater than one. Furthermore, the recoil Ar reconstructed position resolution should be less than 50\,cm, which corresponds to an $L/E$ resolution of less than 2.5\%, and the reconstructed time resolution should be less than 10\,ns to ensure that events are reconstructed in the neutron-free beam-spill time window. 
 With these design considerations, assuming CCM operation at 20\,m from the source for one quarter of the data taking time and at 40\,m for the remaining time, we will be able to make a sensitive search for $\nu_s$'s, and to specifically probe the LSND signal region.

\begin{figure}[htp]
    \centering
    \includegraphics[width=0.48\textwidth]{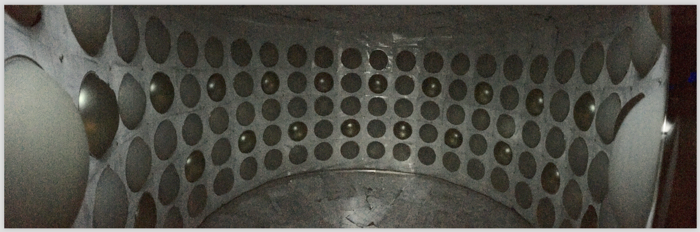}
    \caption{The inside of the CCM120 detector. The 120 inner PMTs are placed around the cylinder barrel, 96 coated, 24 uncoated, and TPB painted reflective foils are also shown.}
    \label{fig:ccm_inside}
\end{figure}
\begin{figure}[htp]
    \centering
    \includegraphics[width=0.48\textwidth]{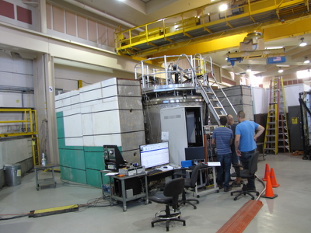}
    \caption{The CCM detector in the ER2 region of the LANSCE Lujan facility with the concrete/poly shielding around it.}
    \label{fig:ccm_environmental_hall}
\end{figure}

\subsection{Physical Description of the Detector}

The experimental layout for the CCM120 detector follows from the design considerations in the previous section. The CAPTAIN detector is an upright cylindrical cryostat 2.58\,m in diameter and 2.25\,m high and holds 10 tons of LAr.  It is equipped with a LAr circulation system with filters to remove impurities including water and oxygen; however, this was not operational for the CCM120 engineering run.  The PMTs operate at positive high voltage, with the photocathodes at ground, so that a single cable provides both the high voltage and readout signal for each PMT.  Finally, the CCM cryostat is designed to be movable for distances between 20\,m to 40\,m from the neutrino source. This provides space for steel and concrete shielding upstream of the detector that will reduce beam neutron background to a sufficiently low level.  The 800\,MeV proton beam strikes the target from above so the neutrinos and DM particles reaching the detector travel at $90^{\mathrm{o}}$ with respect to the beam direction.  

\subsubsection{The phototubes}
There are 120 Hamamatsu R5912-mod2 PMTs (8-inch hemispherical PMTs) mounted on a cylindrical barrel (5~rows $\times$ 24 columns) facing inward toward the 5~ton fiducial volume as shown in Fig.~\ref{fig:ccm_inside}.  In addition, the outer 20\,cm of the cryostat is optically isolated from the inner detector and serves as a veto region for detecting particles that enter or leave the inner volume.  There are 28 PMTs placed in the veto region to reject particles traveling in or out of the detector.  At the bottom endcap, 5 veto PMTS (R5912-mod2) are placed underneath the inner vessel facing the cryostat wall every 72$^{\mathrm{o}}$.  The other 23 PMTs in the veto region are 1-inch Hamamatsu PMTs.  Of the 23 1-inch veto PMTs, 7 are placed above the inner vessel and the remaining 16 are placed around the barrel with a larger concentration on the upstream side of the CCM detector.

The 120 PMTs facing the fiducial volume provides 25\% photocathode coverage.  Of the 120 PMTs, 96 of them have surfaces sandblasted to add scintillation grade Tetraphenyl Butadiene (TPB) wavelength shifter to their surfaces.  The TPB is used to shift the 128\,nm scintillation light into a wavelength ($\sim$425 nm) better matched to the quantum efficiency of the R5912-mod2 PMT.  The other 24 PMTs remain  uncoated.  The combination of coated and uncoated PMTs are used to disentangle the TPB properties when calibrating the detector (see Sec.\,\ref{sec:calibrations}).  The 24 uncoated PMT are located in the 2$^{\mathrm{nd}}$ and 3$^{\mathrm{rd}}$ rows in an alternating pattern (i.e., one in each of the 24 column).  The top, bottom and the region between the 120 PMTs facing the fiducial volume are also covered with reflective foils painted with TPB to improve the light gathering capability of the detector.  The 1-inch~veto PMT's have TPB painted acrylic plates mounted in front of the photocathode.  

Both the 8-inch and 1-inch PMTs use the same cold and warm cables, thus minimizing the timing differences.  However, the 1-inch PMTs have a 42 ns faster response time when compared to the 8-inch PMTs, and this is accounted for in the analysis.

The 8-inch PMTs are prepared for the SBND experiment and are optimized for linearity up to about 50\,PEs.  The single PE pulse height is around 10\,mV, or 40~ADC counts, far above the electronic background ($\sim$~a few ADC counts).  The 1-inch~PMTs in CCM120 are decommissioned from the Mini-CAPTAIN detector. All the PMTs are connected with 8 meters of RG-316 cables inside the cryostat, and 22 meters of LM-195 cables on the outside to be consistent with the SBND requirements.

\subsubsection{Electronics and Data Acquisition}
Eleven CAEN VX1730 boards are used to digitize the signals coming from the CCM PMTs and from surrounding detectors and monitoring devices. The CAEN VX1730 has 14-bit flash ADCs operating at 500 MHz.  The DAQ window is set to 16\,$\mu$s and data from 172 channels are saved for each trigger. With a trigger rate of 22.2\,Hz  (\textit{beam on}) and 2.2\,Hz (\textit{beam off}) about 5.2\,TB of raw data are taken each day. The raw data files, shortly after being created, are passed to a processing script that locates pulses for each PMT and reduces the file sizes by about a factor of 10. These processed files are saved to disk, and each file contains 1000 triggers.

The triggers composing the data stream are beam (20\,Hz), random (1.1\,Hz), and LED (1.1\,Hz). The LED trigger is always 500\,ms after the random trigger. Both the random and the LED are independent of the beam trigger. The CAEN V2495 FPGA board is used to read in three voltage signals and output one trigger signal. The output trigger signal is sent to the boards, and saved in one channel for each board to calibrate time difference between the boards. A copy of the trigger source signal is also saved in the 11$^{\mathrm{th}}$ board with each trigger type posted in a designated channel. Using this technique, it is possible to keep a record of which source caused the trigger.

The 11$^{\mathrm{th}}$ board also records the beam current monitor (BCM) signal that is used to create the beam trigger. The recorded BCM signal is used to remove the time jitter between the beam trigger and the BCM.  Data from six Eljen EJ301 detectors are also saved in the 11$^{\mathrm{th}}$ board. One of the EJ301 scintillator detectors is placed in a neighboring flight path to observe the $\gamma$-flash coming from the protons hitting the target. Data from the neighboring EJ301 detector is used to determine the earliest neutrinos arriving in the CCM detector.


 \section{Beam and Particle Sources \label{sec:particlesources}}

\subsection{Beam Characteristics}
The triangular beam delivered to the tungsten target in the Lujan Center (Fig.~\ref{fig:ProtonBunchTime}) is determined by the orbit time of the 800-MeV protons in the storage ring (PSR).  The ion source’s large beam structure is sliced into 1750 segments that can be injected turn-by-turn into the ring’s 358-ns time-of-flight window. Approximately 80\,ns is needed for extraction of the beam, leaving the 280-ns width observed. Each slice is referred to as a ‘mini-pulse.' The PSR uses a buncher to maintain beam stability.  If the single mini-pulse injection were used, it is possible for the harmonic buncher to narrow the 280\,ns beam width to less than 100\,ns.  However, with full production, injected mini-pulses near the end of accumulation maintain the original width, giving the appearance of a triangular profile instead of the normal distribution of a bunched beam.  Twenty macro-pulses are accumulated in the PSR and delivered to the tungsten target every second, and each macro-pulse contains  $3.1 \times 10^{13}$ protons. 


\subsection{$\pi^+$ Production}
$\pi^+$ decay is the main source of neutrinos produced in the Lujan target. The majority of $\pi^+$ are stopped by nuclear interactions before decaying nearly at rest, providing the characteristic mono-energetic prompt neutrino source expected by coherent neutrino nucleus scattering experiments. 

The number of $\pi^+$ produced is simulated with MCNP6.2 \cite{MCNP} and the LANSCE Lujan Center Mark-III Target System target card.  From the results of the simulation, 0.05717 $\pi^+$ are produced per proton on target (POT); however, only 0.04586 $\pi^+$ per POT decay. The remaining, approximately 20\%, are absorbed by nuclear interactions before they can decay, consistent with previous measurements \cite{Allen:1989dt}. MCNP6.2 imposes a cutoff in tracking pion energies of 1 keV, at which point they are automatically decayed. For the Lujan target simulation, only $5\times10^{-4}$ $\pi^+$ per POT (1\% of all decays) are found to decay in flight before reaching the low energy cutoff.

\subsection{$\pi^0$ production}

\begin{figure}[htp]
    \centering
    \subfloat[$\cos\theta$\label{fig:pi0_angle}]{\includegraphics[width=0.40\textwidth]{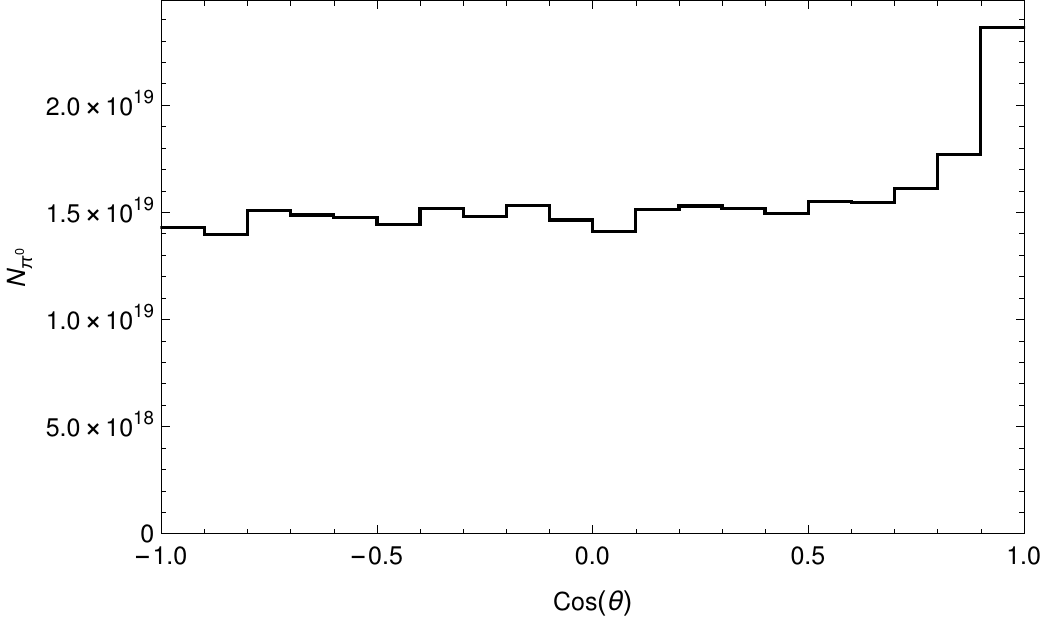}}

    \subfloat[$p_{\pi^0}$\label{fig:pi0_momentum}]{\includegraphics[width=0.40\textwidth]{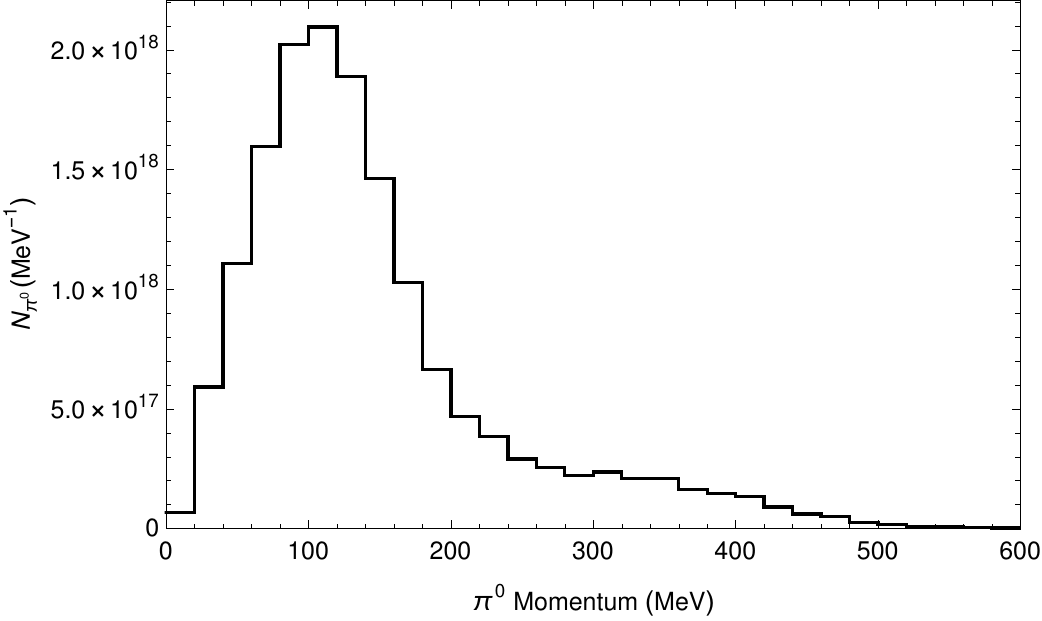}}
    \caption{The $\pi^0$ (a) angular and (b) momentum distributions produced at the Lujan target assuming POT=$2.71\times10^{21}$. The angular production is mostly isotropic, with a significant peak in the production parallel to the beam. The momentum distribution peaks between 100 and 120\,MeV, with a mean momentum of 146\,MeV. The CCM120 detector is $90^{\circ}$ with respect to the beam direction.}
    \label{fig:pi0_dists}
\end{figure}

$\pi^0$s are produced in copious numbers in proton-target collisions at Lujan; however, a simulation must be used to determine the number and distribution of $\pi^0$s produced. The $\pi^0$ has a lifetime of $8.5\times10^{-8}$\,ns \cite{Nu_mass}, and will therefore decay in flight rather than slow down and stop through nuclear interactions. The $\pi^0$ production is simulated using MCNP6.2 \cite{MCNP} and the LANSCE Lujan Center Mark-III Target System target card. The total number of $\pi^0$s, $N_{\pi^0}$, scales linearly with the number of Protons on Target, and we find that $N_{\pi^0}=0.115 \times \mathrm{POT}$. The distribution is close to isotropic over most angles but with a noticeable enhancement in the forward direction, as shown in Fig.\,\ref{fig:pi0_angle}, and peaks between 100 and 120 MeV as shown in Fig.\,\ref{fig:pi0_momentum}.

\subsection{LDM production}

We consider a scalar DM candidate coupled to the SM through the vector portal as a benchmark scenario for MeV-scale DM \cite{deNiverville:2012ij},
\begin{eqnarray}
\label{eq:lagrangian}
  {\cal L} = & -\frac{1}{4} V^2_{\mu \nu} - \frac{1}{2} m_{V}^2 V^2_\mu + \varepsilon V_\nu\partial_\mu F_{\mu \nu} + |(\partial_\mu - g_D V_\mu)\chi|^2 \nonumber\\
  & -m_\chi^2 |\chi|^2 + {\cal L}_{h^\prime}.
\end{eqnarray}
The dark photon, $V$, can be produced on- or off-shell through mixing with the SM photon, and then decay into a pair of DM particles $(\chi \bar{\chi})$. For the relatively low energies of the proton beam at Lujan, the greatest source of DM is the radioactive decay of the $\pi^0$. In the on-shell approximation, the branching ratio is given by \cite{Kahn:2014sra}
\begin{eqnarray}
\label{eq:mesondecay}
\mathrm{Br}(\pi^0 \to \gamma \chi \bar{\chi}) \, = \, & 2 \varepsilon^2 \mathrm{Br}(\pi^0\to \gamma \gamma) \left(1-\frac{m_{V}^2}{m_{\pi^0}^2}\right)^3 \nonumber\\
& \times \mathrm{Br}(V \to \chi \bar{\chi}),
\end{eqnarray}
where $\mathrm{Br}(\pi^0 \to \gamma \gamma) = 0.988$ \cite{Nu_mass}. The full expression including off-shell production is available in Ref.\,\cite{Kahn:2014sra}. Radiative $\pi^-$ capture is also possible, but expected to be heavily suppressed for heavy elements like the tungsten used in the Lujan target\,\cite{Pospelov:2017kep}.

\begin{figure}[htp]
    \centering
    \subfloat[$m_\chi=10$\,MeV]{\includegraphics[width=0.40\textwidth]{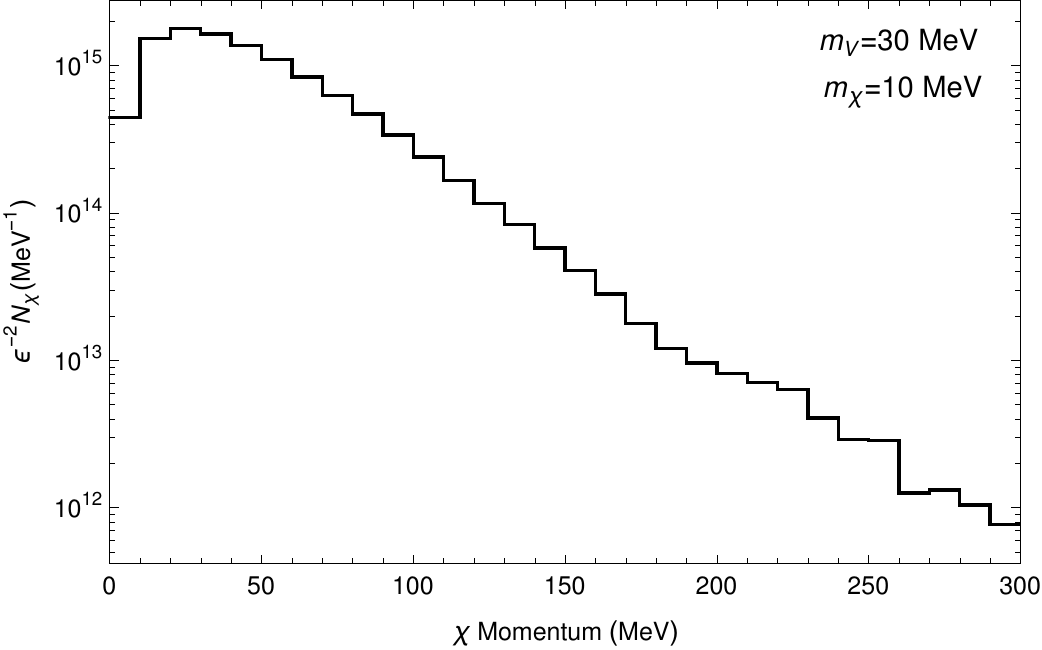}}
    
    \subfloat[$m_\chi=50$\,MeV]{\includegraphics[width=0.40\textwidth]{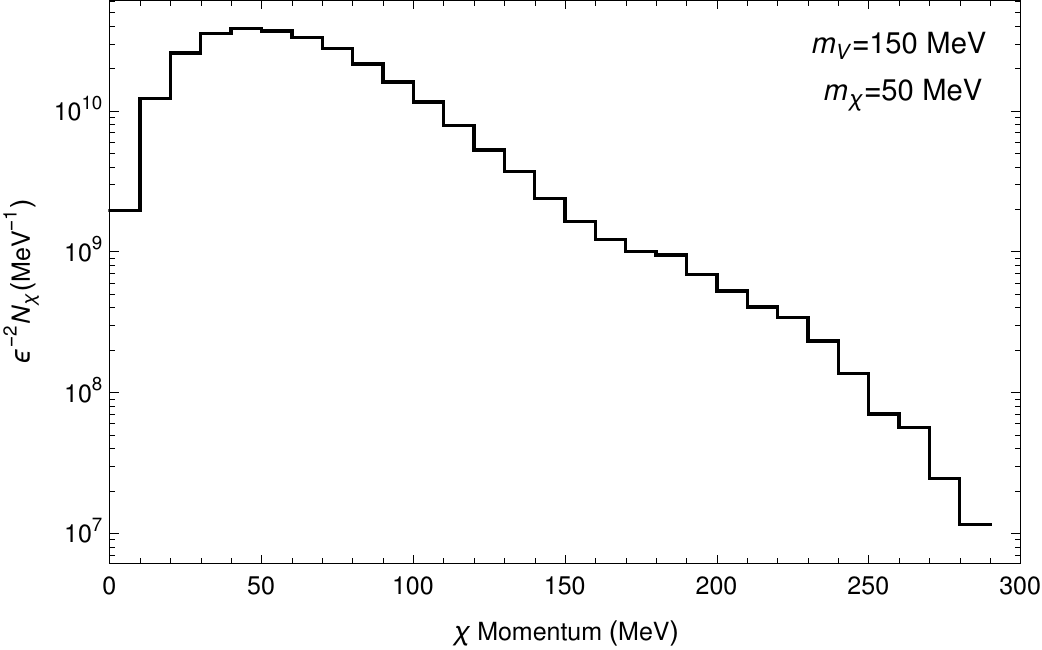}}
     \caption{The $\chi$ momentum distribution expected by the CCM detector, assuming POT=$2.71\times10^{21}$ and (a) $m_\chi=10$\,MeV and (b) $m_\chi=50$\,MeV, where the dependence on $\varepsilon$ has been removed.}
    \label{fig:DM10_50mev}
\end{figure}

The decay of the $\pi^0$ into DM particles is simulated using the BdNMC simulation\,\cite{deNiverville:2016rqh}. The generated DM momentum distributions incident on the CCM detector are shown in Fig.\,\ref{fig:DM10_50mev} for two values of $m_\chi$.

The vector boson $V$ could also couple to baryon current $J_\mu^B \equiv \frac{1}{3}\sum_i \bar q_i \gamma_\mu q_i$~\cite{Batell:2014yra,Coloma:2015pih,Dror:2017ehi,Berlin:2018bsc,Boyarsky:2021moj}. The additional coupling to quarks can be written as 
\begin{equation}
    \mathcal{L} \supset g_{_B} V_B^\mu J_\mu^B,
\end{equation}
where $g_{_B}$ is the $U(1)_B$ gauge coupling strength. This scenario posits a leptophobic DM candidate by assuming that the kinetic mixing $\varepsilon \ll \alpha_{_B}$, where $\alpha_{_B} \equiv g_{_B}^2/\left(4\pi\right)$. In order to remain consistent with the formalism of Ref. \cite{Boyarsky:2021moj}, the coupling between the DM $\chi$ and $V_B$ is called $g_\chi$ with $\alpha_\chi \equiv g_\chi^2/\left(4\pi\right)$, where $\alpha_{_\chi}$ in the leptophobic scenario is equivalent to $\alpha_{_D}$. Note that $\alpha_{_B}$ and $\alpha_\chi$ can take on very different values, and this work will fix $\alpha_\chi = 0.5$ while keeping $\alpha_{_B}$ as a free parameter. 

The coupling to the baryon current can induce an effective kinetic mixing of $\varepsilon \sim e g_{_B}/\left(16 \pi^2\right)$, which will not affect the phenomenology of the model at CCM but is important for the relic density calculation.

As in the case of kinetic mixing, LDM will be produced through $\pi^0$ decays~\cite{Batell:2014yra}, 
\begin{equation}
    \mathrm{Br}(\pi^0 \to \gamma V_B) = 2\left(\frac{g_{_B}}{e} - \varepsilon\right)^2\left(1-\frac{m_V^2}{m_{\pi^0}^2}\right)^3 \mathrm{Br}(\pi^0\to\gamma\gamma).
\end{equation}
There can be interference between the kinetic mixing and the baryonic coupling, suppressing $\pi^0 \to \gamma V_B$ for $\varepsilon \approx g_{_B}/e$, though this work will not explore this region of the parameter space.

\section{Data Analysis \label{sec:dataanalysis}}
\subsection{Pulse Finding and Event Building}
\begin{figure}[ht]
    \centering
    \includegraphics[width=0.48\textwidth]{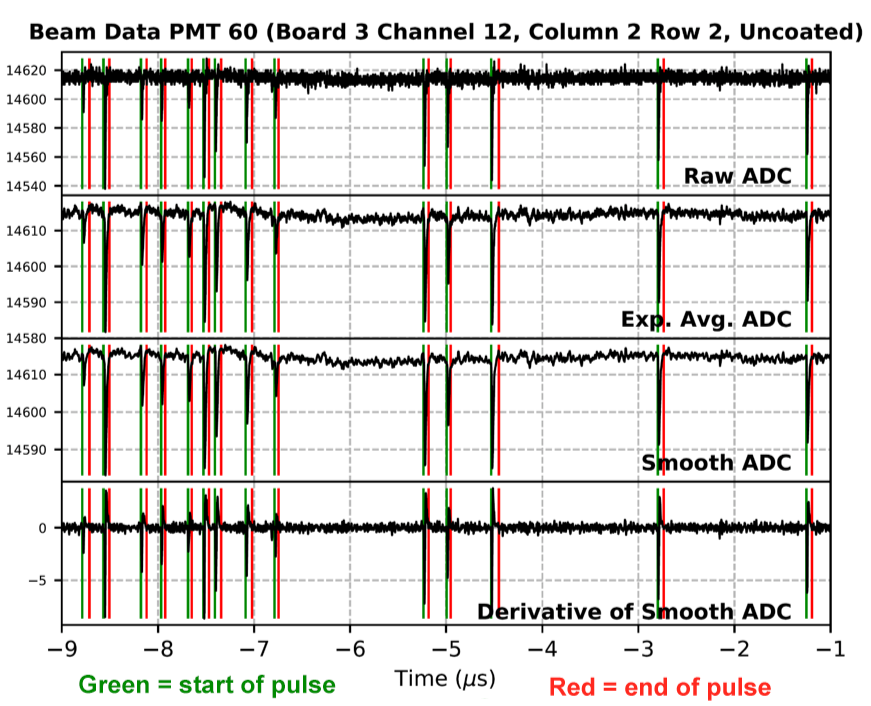}
    \caption{The process for finding pulses from a PMT waveform. (top row) The raw waveform coming from the CAEN digitizers. (second row) Exponential average of the raw waveform. (third row) Running average or smoothed waveform of the exponential average. (bottom row) The derivative of the waveform after all smoothing. The start (green) and stop (red) of each pulse, after a 5 ADC integral threshold is applied, are marked to point out where single photoelectrons (SPEs) are observed.  Note how the noise is reduced as subsequent smoothing techniques are applied.}
    \label{fig:pulse_finding}
\end{figure}

The first step in the event building process is to find individual pulses using the waveforms from each PMT. The pulses are grouped together to create an accumulated waveform and used to identify physics events in the detector. The process of finding individual pulses is shown in Fig.~\ref{fig:pulse_finding}.
The waveforms from each PMT are first smoothed with an exponential average and then a running average is applied to remove high frequency and white noise. Following this, the derivative of the waveform is calculated. Pulses are found using the derivative technique to minimize sensitivity to the known overshoot problem with the PMTs~\cite{Kaptanoglu:2017jxo}.

Since the pulses can be approximated as inverted Gaussian distributions, a pulse starts when the derivative goes negative and ends after the derivative goes to zero after going positive first. The only selection criteria applied to the pulses before calibration are to require them to have a minimum duration of 10\,ns, and also have an absolute value of the derivative that crosses the threshold. The absolute integral of the derivative, $I_D$, calculated in units of ADC, is used as a proxy for the charge.

From the LED calibrations, a threshold value of 5\,ADC on $I_D$ is determined for each channel based off the measured noise from the electronics when no HV is supplied to the PMTs. A length cut of 20\,ns on the pulse removes a lot of the leftover noise after the HV was turned on.

\begin{figure}[htp]
\centering
\subfloat[Longer DAQ Window\label{fig:accum_wf_1}]{\includegraphics[width=0.48\textwidth]{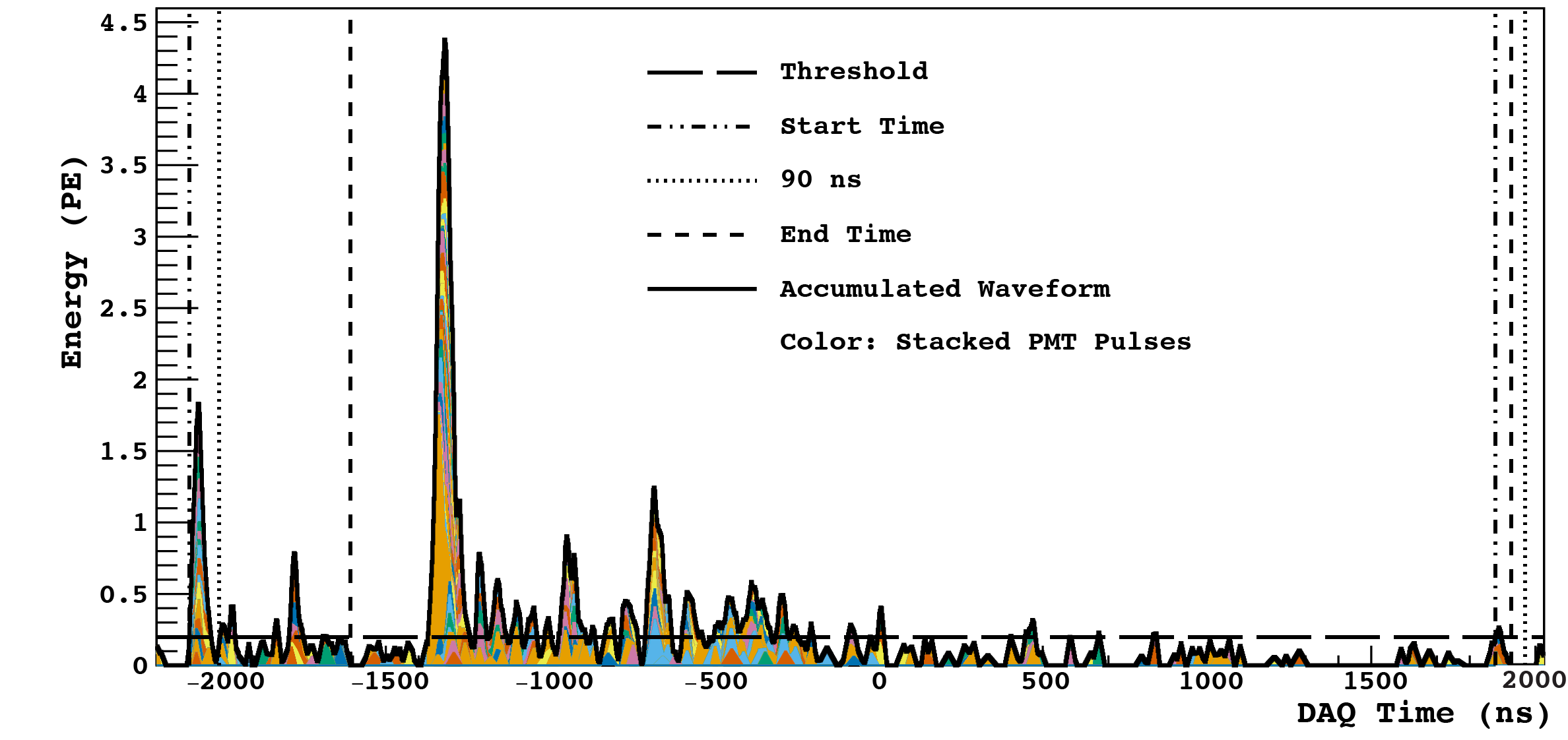}}\\
\subfloat[Shorter DAQ Window\label{fig:accum_wf_2}]{\includegraphics[width=0.48\textwidth]{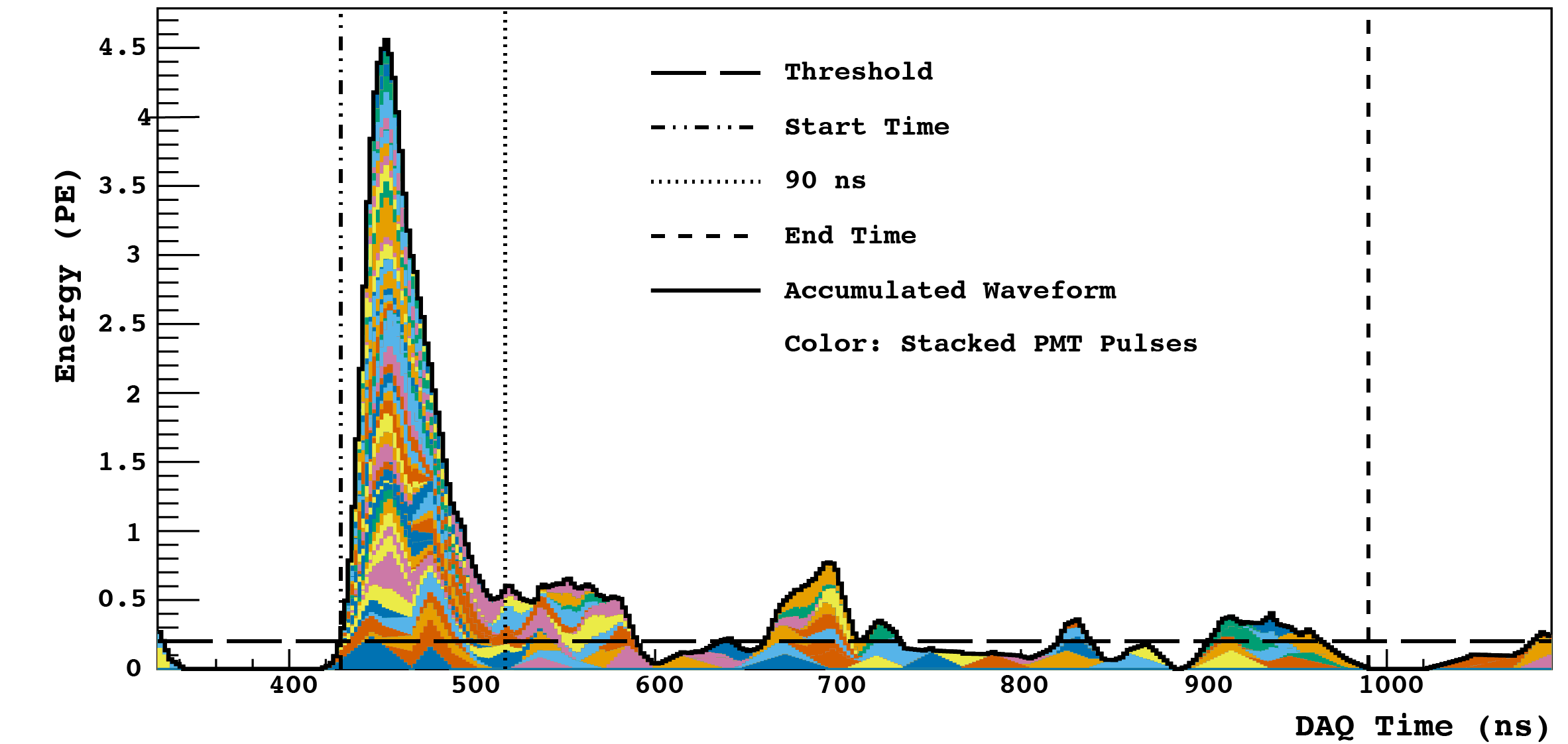}}
\caption{Examples of an accumulated waveform where each PMT is represented by a different color (9 colors for 120 PMTs). The pulses from the PMTs are represented by a triangle with an area equal to the integral of the pulse. The threshold is indicated by a long dashed line. The start, 90\,ns, and end of an event are represented by dashed-dot-dot, dot, and dashed lines respectively. Only events that passed source calibration cuts have start, 90\,ns, and end lines.}
\label{fig:accum_wf}
\end{figure}

To build events, pulses from the individual PMTs are accumulated into one waveform as shown in Fig.~\ref{fig:accum_wf}.
A triangle is used to represent a given pulse, where the length of the triangle is equal to the length of the pulse and the area of the triangle is equal to $I_D$. The accumulated waveform used to find events is then the sum of all the triangles.

Beam-related backgrounds associated with the LDM search are discussed in more detail in Sec.\,\ref{sec:dmCuts}.  For now, it will suffice to say that because of a large amount of beam-in-time beam-related backgrounds, a long time integral cannot be used for the DM analysis. Therefore, to build candidate events each event starts when a given 2\,ns bin crosses a threshold of 0.2\,PE and ends when there is 20\,ns of nothing seen from the 120 8'' PMTs. The reconstructed energy of the event is then represented as the integral of PE observed in the first 90\,ns or length of the event, whichever one is shorter. Sec.\,\ref{sec:dmCuts} will describe further data selection criteria used in the DM analysis.

\subsection{\label{sec:calibrations}Calibrations and the Optical Model}
Three calibration techniques are used to explore and characterize the overall response of the CCM detector.  LEDs are used to calibrate the photoelectron (PE) response of the PMTs. \iso{57}{Co} and \iso{22}{Na} sources are used to calibrate the energy scale of the detector at 126\,keV and 2.2\,MeV respectively. A laser, operating between two different frequencies, is used to understand the absorption, wavelength shift, and reflective properties of the detector. The following is a detailed description of each calibration system.

    The calibration data are used in the construction of a simulation program called the ``Optical Model.''  A thorough understanding of the light propagation inside the CCM detector is required for the data analysis, so a monte carlo program based on GEANT4 \cite{Agostinelli:2003g4} is used to simulate the light propagation and quantify the detector response.  The GEANT4 program incorporates optical processes such as scintillation, Rayleigh scattering, optical absorption and reflection, and wavelength shifting (WLS). It is also equipped to give the user control of the variables needed to construct and customize the Optical Model. 

Calibration data is used to customize the detector response in the Optical Model.  More precisely, the data are used to simulate the contaminated argon, leeched TPB, and reflective foils inside the detector. Three sets of calibration data were collected during the engineering run and they are described below. 

\subsubsection{\iso{57}{Co} and \iso{22}{Na} Sources\label{sec:radioactiveSources}}
The first data set used in the detector calibrations comes from the source calibrations (\iso{57}{Co} and \iso{22}{Na}) which recorded the light output from the known $\gamma$-ray spectra.  The light was tracked from the source to the PMTs to determine the number of PEs produced in each PMT. A 3.05\,m rod was used to insert a radioactive source along the vertical axis of the chamber, and positioned it from the top to the center of the detector.  The source was encapsulated in a stainless steel capsule at the end of a 0.30\,m bayonet to maximize solid angle acceptance. 

Two sources were chosen to calibrate the detector: (i)  a 10\,$\mu$Ci \iso{57}{Co} source produced two $\gamma$-rays $\sim$126\,keV that escaped the stainless steel capsule and (ii) a 3\,$\mu$Ci \iso{22}{Na} source produced back-to-back 511\,keV $\gamma$-rays (due to positron-electron annihilation) simultaneous with a 1.27\,MeV $\gamma$-ray. When the sources were placed in the detector their decay products were completely contained, thus providing an energy calibration between 126\,keV and 2.2\,MeV. 

The DM search described in Sec.\,\ref{sec:dmCuts} is a prompt only analysis, and therefore the source calibration is done for prompt light only. The prompt time region, first 90\,ns of the event, corresponds to about 30\% (70\%) of the light coming from electromagnetic (nuclear recoil) events.
\begin{figure}[htp]
    \centering
    \includegraphics[width=0.50\textwidth]{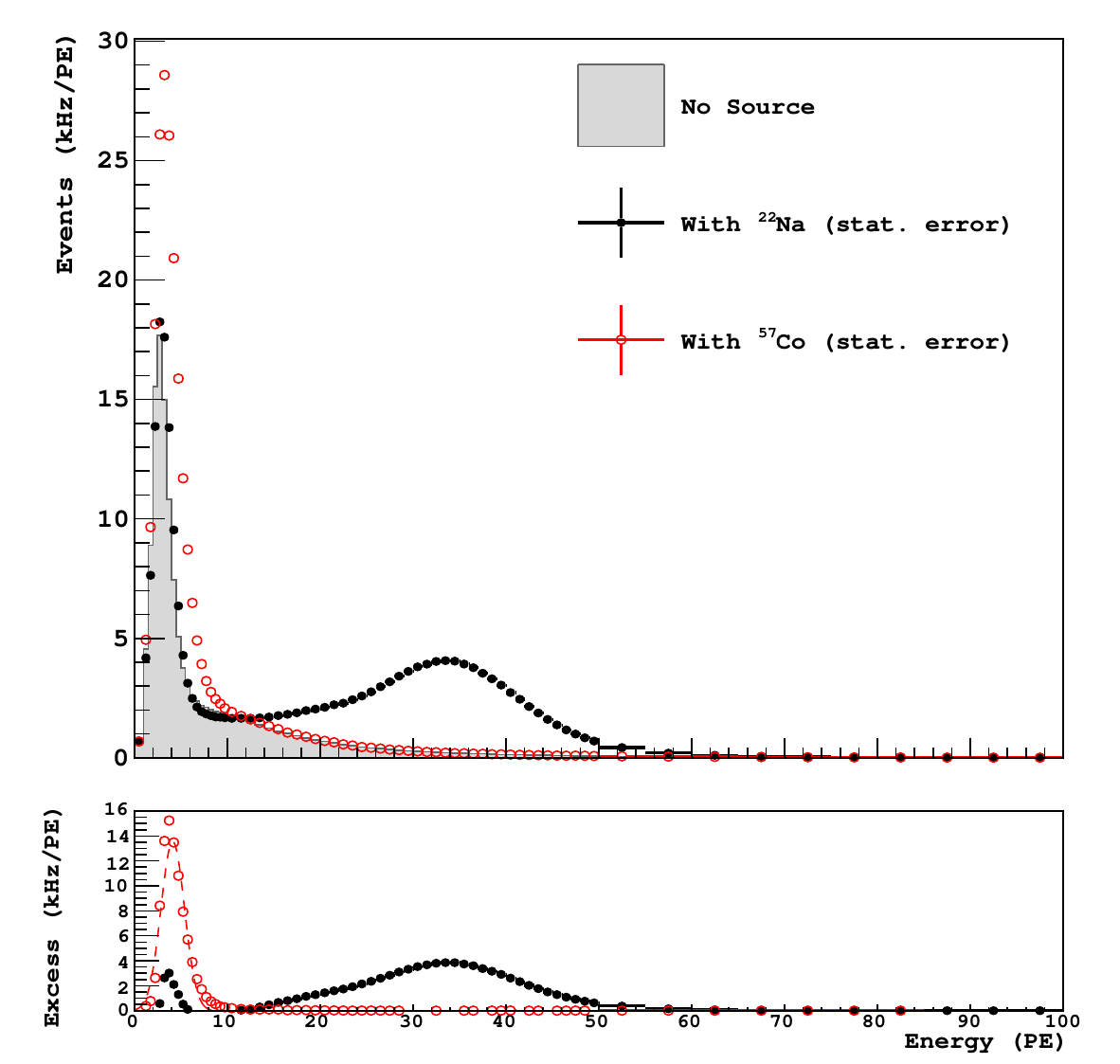}
    \caption{A comparison between \textit{with} and \textit{without} the 3\,$\mu$Ci \iso{22}{Na} or the 10\,$\mu$Ci \iso{57}{Co} is shown for sources in the center of the detector. (lower) The excess event rate for each source.}
    \label{fig:na22Calib}
\end{figure}
The energy calibration is determined from the \iso{22}{Na} source, which is more robust, and found to be $15.1\pm4.0$\,PEs/MeVee (electromagnetic equivalent).  This corresponds to a 4.1\,MeVnr (nuclear recoil) source, including a quenching factor of 27.2\% for the corresponding energy region.  Unfortunately, the true peak for \iso{57}{Co} is less than 2 PE and not observable above the noise.  However, we assumed a linear response for the energy calibration using the \iso{22}{Na} source and trusted it for energies $>$200~keV (i.e., energies greater than the \iso{57}{Co} decay).

The uncertainty obtained from the quenching factor is determined by taking the world measurements and finding the energy independent mean and sigma~\cite{Akimov:2020pdx}. Unlike what is shown in Ref.~\cite{Akimov:2020pdx}, where a linear slope is determined, an energy independent average is needed because the nuclear recoil events from DM coherent scattering extends up to \begin{math}\mathcal{O}\end{math}(1\,MeV) and the current quenching factor data goes up to around 200\,keV. Therefore, the quenching factor used for this analysis is $0.272\pm0.035$, which corresponds to a fractional error of 12.0\% in reconstructed DM events.

\subsubsection{LED}
Blue LEDs ($\lambda$=465\,nm) are placed at the top and bottom of the detector and used to determine the SPE response of the PMTs.  The LEDs are pulsed at 1.1\,Hz to calibrate the detector over the course of the run. The pulse duration for each LED is 10\,ns and each PMT is tuned to a single PE response. The time region in the DAQ before the LED light is on is used for background subtraction for the PMTs.

On average the single PE response of the PMTs in liquid argon is $9.26\pm3.76$\,ADC units with a minimum of 2.26 and a maximum of 26.42. This is about half of what is measured in warm conditions and consistent with Ref.\,\cite{Babicz:2018svg}. For CCM200, shorter cables and updated PMT bases will be used to improve the single PE response.

\subsubsection{Laser Calibration}
A custom Q-switched diode pumped Q1B-10 laser made by Quantum Light Instruments is used as a light source for PMT calibration. Leveraging the use of 2$^{\mathrm{nd}}$, 4$^{\mathrm{th}}$, and 5$^{\mathrm{th}}$ harmonic generators with an initial beam at 1064\,nm, this laser emits light at 213\,nm and 532\,nm. The maximum output energy is 1\,mJ and 0.5\,mJ respectively with a pulse length $<$\,8\,ns. The laser is equipped with remote control of (i) the input pump current as well as (ii) the mechanical scroll-step variable attenuator to reduce total beam transmission. The variable attenuator has a 0.2\% transmission adjustment accuracy.


\subsubsection{Optical Model}


The parameters used in the Optical Model are derived from the calibration data and they accurately simulate the contaminated argon, leeched TPB, and reflective foils inside the detector. The parameters depend mostly on two sets of calibration data collected during the engineering run.  Other parameters are determined from ratios of PMT responses between various rows (5 ratios in all) and are used to determine the attenuation length.  The PMT responses depend on the wavelength and the height of the laser diffuser.  Five different heights and five ratios between rows of PMTs produced 25 measurements.  Adding in a ratio between \textit{coated} and \textit{uncoated} PMTs resulted in 50 data points used in the $\chi^2$ analysis of the data.

The simulations explored a large range of parameters in an attempt to find which ones significantly affected the $\chi^2$. Some were quickly eliminated; in one case, the optical absorption of visible light ($>$ 400 nm) did not alter the $\chi^2$ value unless the absorption length was reduced to below 1000 cm, less than half the value found in the literature (2000 cm)\,\cite{Rust64}. In another case, attempting to alter the reflectivity of the teflon foils from values of 95\% visible and 10\% UV caused the $\chi^2$ to rapidly rise; therefore, the reflectivity was not adjusted during the fit. 

The following is a list of variables considered in the $\chi^2$ minimization and they are divided into three groups. The first group includes a few variables not varied during the optimization and the reason(s) for not including them. The second group are variables determined by the source row ratios, and the third group are the variables determined by the laser row ratios.
\begin{center}
    \textit{Parameters Not Included as Variables}
\end{center}
\textit{Absorption length of visible light in liquid argon:} the value found in the literature is 2000 cm, and simulations found that both laser and sodium $\chi^2$ remained constant between 1400 and 3400 cm.  \cite{Rust64}\\

\textit{Reflectivity of the mylar foils in UV and visible wavelengths:} this is the optical reflectivity of the mylar at the wavelengths considered to be important. These values are 95\% (visible) and 10\% (UV) from the literature, and alteration of these values caused the $\chi^2$ to quickly rise. Thus, the literature values were accepted as correct and further tuning of this parameter was not pursued. \\

\textit{TPB efficiency of the PMTs:} this is the efficiency of the TPB evaporatively coated onto the PMTs. Varying this value has minimal effect on either Laser or Source $\chi^2$ over a reasonably large spectrum so long as the efficiency is kept over 85\%. A literature spectrum is used to map out the efficiency based on a 90\% efficiency at 128 nm \cite{Gehman_2011}. \\

\begin{center}
\textit{Source Determined Variables}
\end{center}
\textit{Abs100:} The absorption length of $<$\,200 nm light. This parameter is the absorption length covering the argon scintillation spectrum \cite{Babicz:2018svg} \cite{Calvo2016}. As such it is the most important variable determined from the source calibration $\chi^2$. This variable also measures the contamination, especially oxygen and water. \\

\textit{R1Clouding, R1Radius:} absorption length modifier and radius for the top of the detector. In performing simulations, it was found that the $\chi^2$ could be improved by increasing the contamination near the top of the detector, specifically towards the edges where there is more mass to leak the contaminant gases. Both variables were initially justified by the significant reductions of source $\chi^2$, though later searches found that the lower mass of common contaminants (oxygen, water, ozone) and reduced boil off as compared to argon did imply a higher concentration near the top. \cite{Corbett1956}\\

\textit{FoilEff:} TPB efficiency of the foils: this is the efficiency of the TPB foils, using a model where the conversion happens over the full thickness and some UV light is allowed to pass through and strike the reflective foils on the other side.  The value listed is specifically for 213\,nm light, which was used as an anchor by which to produce the TPB efficiency spectrum of painted TPB found in the literature.~\cite{Ignarra_2013}

\begin{center}
\textit{Laser Determined Variables}
\end{center}
\textit{Conewide, Conehigh:} LED cone radius and height. Due to the domination of the laser row ratios by the bottom geometry, especially near the center, a foil shape to mimic the location of the calibration LED was included. The two parameters were used to determine an approximation that would best simulate the shape and foil distortions of the actual LED.\\

\textit{Abs200:} Absorption length of 200-300~nm light. This variable was chosen as the absorption length that affects the UV Laser (213\,nm) and little else as the argon scintillation spectrum and the TPB emission spectrum are both outside these wavelengths. \\

\textit{Abs300:} The absorption length of 300-350~nm light. According to literature, this value would be affected by ozone contamination in the liquid argon. Ozone was one of the few contaminants found that would modify the absorption length of 213\,nm laser light \cite{Marcus2013}. This variable was included to accompany the Abs200 variable above.\\

\textit{TopDivider:} TPB thickness on the top foils.  It was found that reducing the thickness and thus the efficiency of the top foils improved the Laser $\chi^2$. This was born out by simulations, and only the top thickness was included in the final set of parameters. \\

\textit{Unsmooth:} segmented bottom foils: We used a single variable to compensate for the less than smooth placement of foils on the bottom of the detector. This parameter describes the randomization of reflected photons away from the simulated reflection off a perfect plane in arbitrary units. \\

The final set of 10 parameters and their uncertainties are listed in Table \ref{bestFitTable}. A covariance matrix was formed from these variables and used to produce correlated ``throws'' of the Optical Model for the purposes of determining the error in the signal.  It is clear that the measured 128\,nm scintillation light ({\em Abs100}) is about a factor of 20 less than the expected attenuation length in clean argon. It was determined from this analysis  that O$_2$ and H$_2$O impurities at the level of 10\,ppm in the argon reduced the 128\,nm scintillation light attenuation to the levels observed, which is consistent with the levels of impurities in LAr delivered from the gas plant.  This was one of the main lessons learned from CCM120 that will be fixed for CCM200 with the commissioning of  the LAr filtration and re-circulation system, which will reduce contamination below 100\,ppb necessary to increase light output back to nominal levels and achieve 10\,keV thresholds.

\begin{table}[h]
\caption{Table of parameters that best optimize the detector response using the Optical Model}
\begin{ruledtabular}
\label{bestFitTable}
\begin{tabular}{l c c c} 
Paramter &  Value &  Error & Units \\
\hline
Abs100 & 55.95 & 6.92 & cm\\
R1Clouding & 12.23 & 5.92 & percent (\%)\\
R1Radius & 64.05 & 11.08 & cm\\
FoilEff & 45.55 & 7.97 & percent (\%)\\
Conewide & 7.555 & 1.488 & cm\\ 
Conehigh & 4.457 & 0.567 & cm \\
Abs200 & 37.55 & 18.71 & cm \\
Abs300 & 1310 & 172 & cm \\
TopDivider & 26.12 & 14.17 & divider \\ 
Unsmooth & 2.922 & 0.480 & arbitrary \\  [1ex] 
\end{tabular}
\end{ruledtabular}
\end{table}

\subsection{Simulating DM Coherent Scattering \label{sec:simDMResponse}}

\begin{figure}[htp]
    \centering
    \subfloat[Efficiency\label{fig:eff}]{\includegraphics[width=0.35\textwidth]{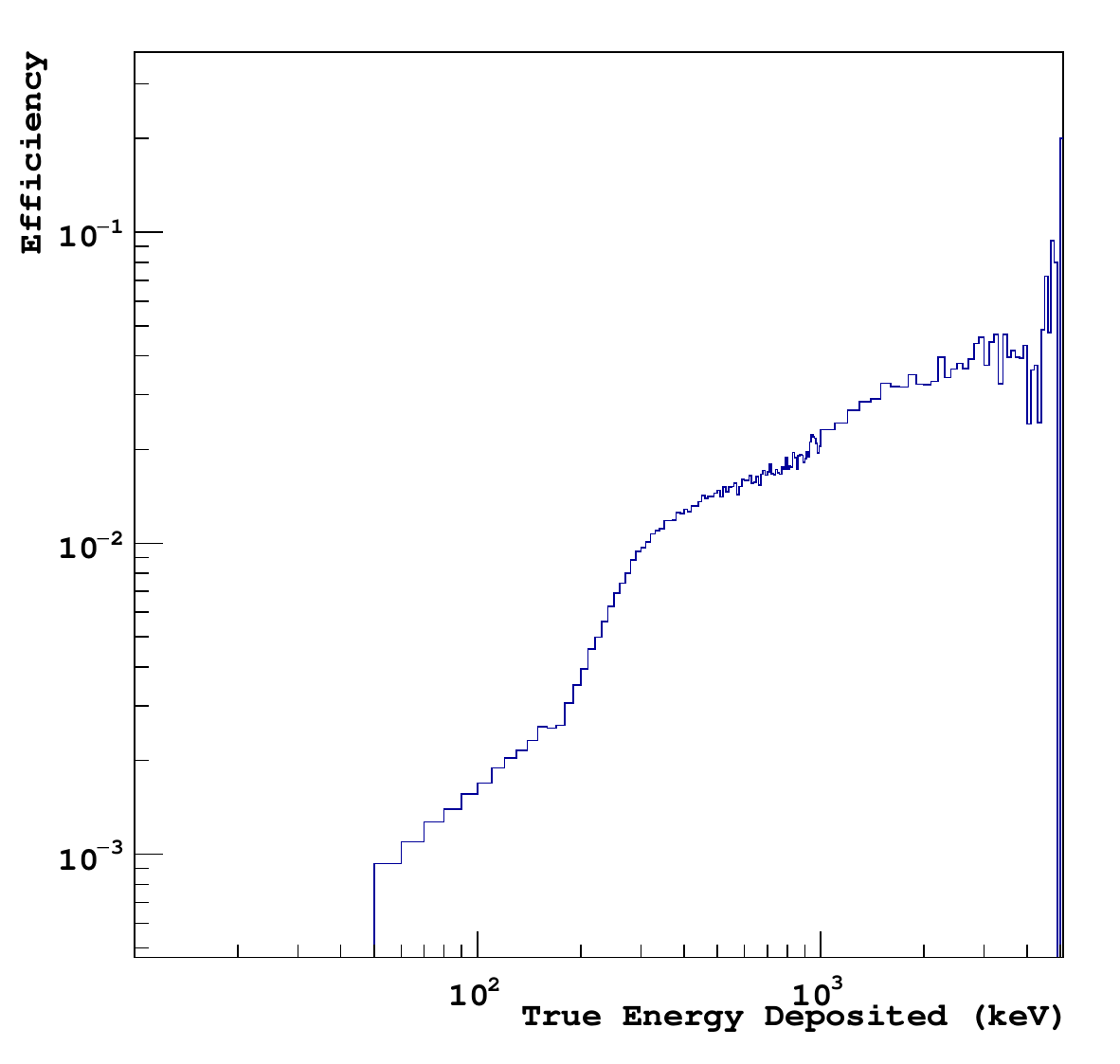}}\\
    \subfloat[Response Matrix\label{fig:responseMatrix}]{\includegraphics[width=0.35\textwidth]{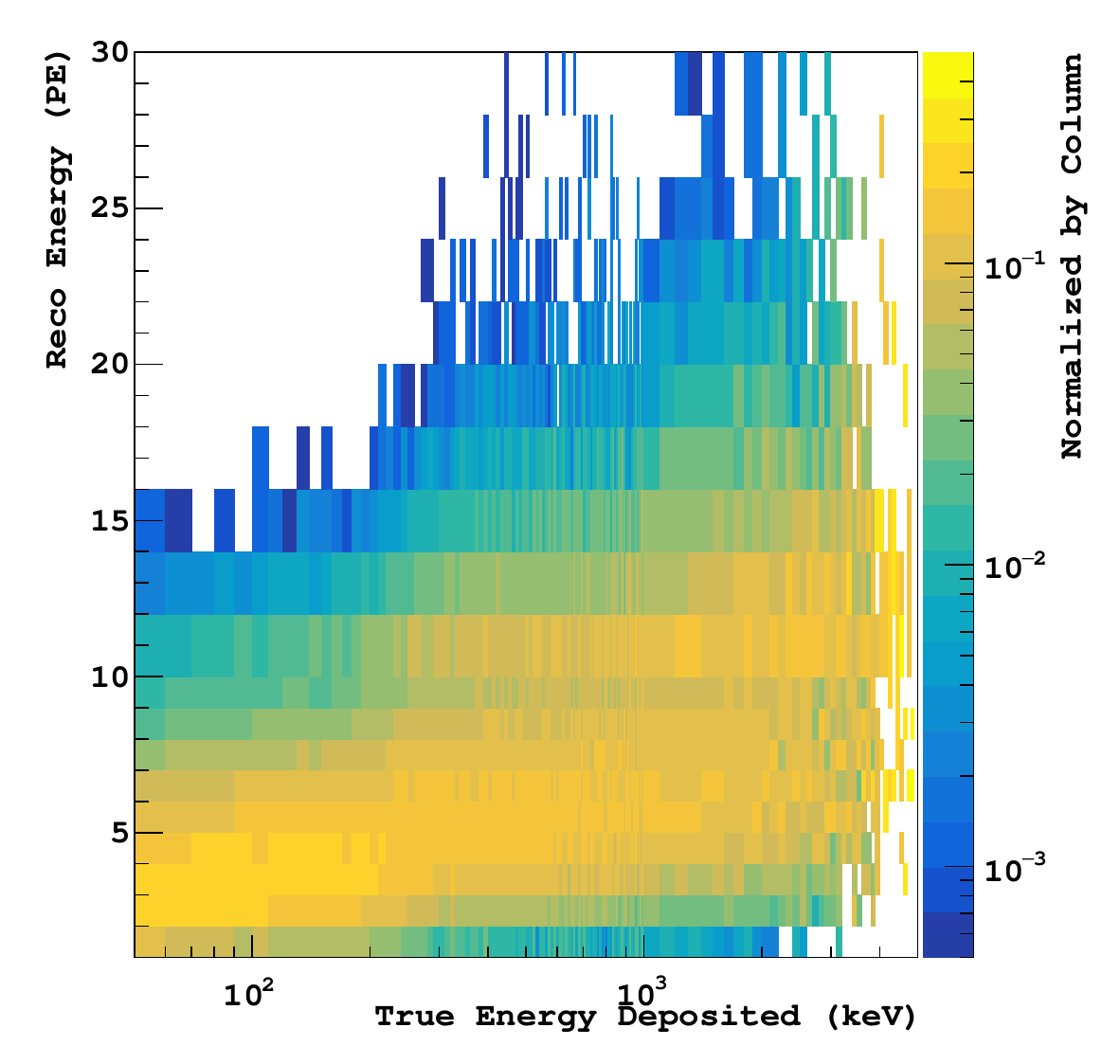}}\\
    \caption{The detector response to a nuclear recoil where (upper) is the total efficiency as a function of true recoil energy, and (lower) is the smearing matrix from true recoil energy to reconstructed energy where each column is normalized so the sum in the color equals one.  Both distributions were constructed after DM selection criteria were applied.}
    \label{fig:nucleonResponse}
\end{figure}

A simulation of nuclear recoil argon events with recoil energies up to 5\,MeV was used to map out the response matrix of CCM120. The response is summarised in Fig.~\ref{fig:nucleonResponse}.

Because of the amount of contamination in the liquid argon, the highest PE/keVnr (nr=nuclear recoil) occurs closest to the TPB where the light is converted to a wavelength with much longer attenuation. The center of the detector, where the source calibrations are located, has the least efficiency. The total efficiency and response matrix of the detector are calculated with the distance-to-wall dependence included. The efficiency obtained is shown in Fig.~\ref{fig:eff} where the efficiency for observing a nuclear recoil event with recoil energy less than 100\,keVnr and 1\,MeVnr is less than 2\% and 20\% respectively.

The mapping from true nuclear recoil energy deposited to reconstructed energy is shown in Fig.~\ref{fig:responseMatrix} where each true energy column has been normalized so the column sums to one. Because of the amount of contamination in the liquid argon there is almost no energy resolution below 7 PEs. 

\begin{figure}[htp]
    \centering
    \includegraphics[width=0.48\textwidth]{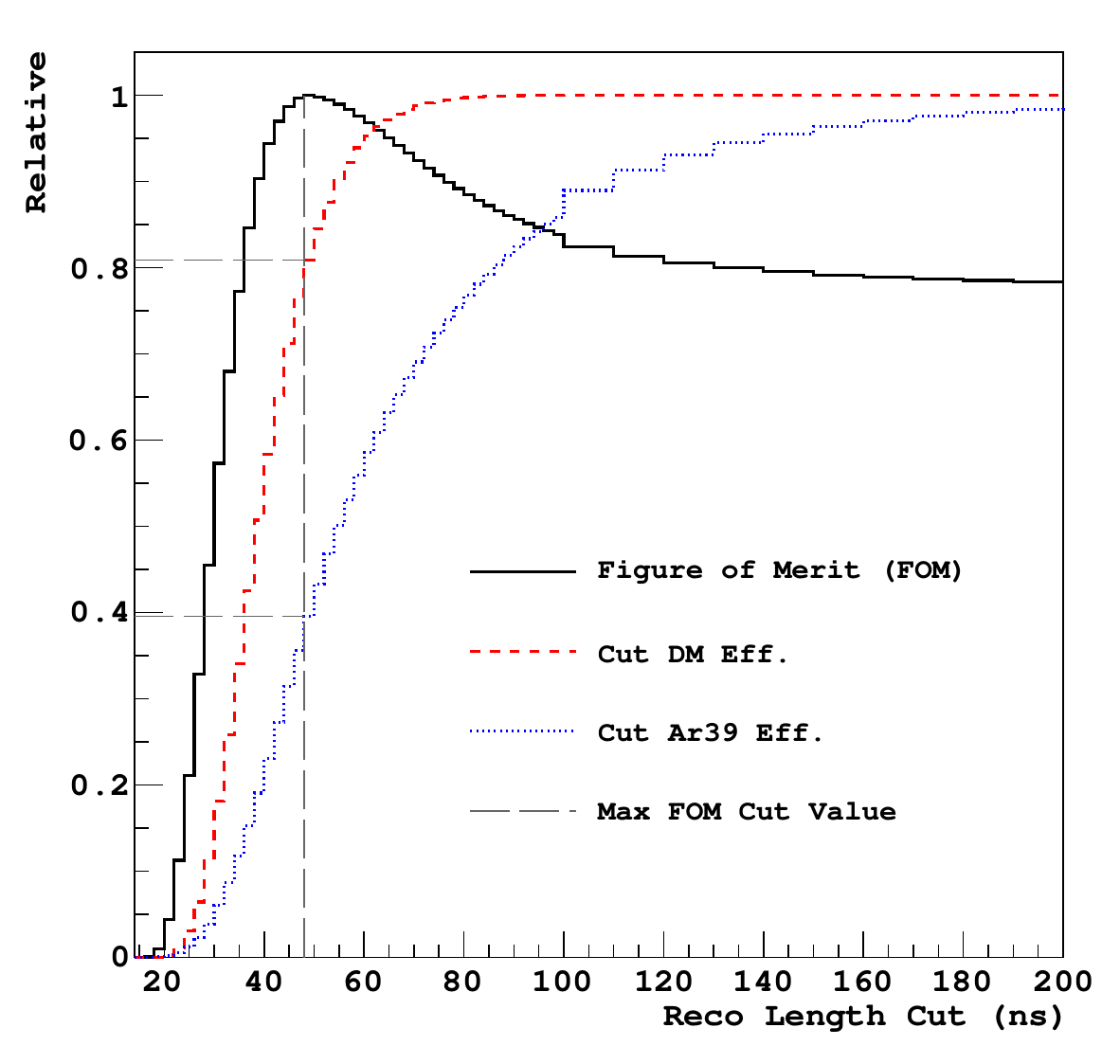}
    \caption{The solid line shows the figure of merit calculation for the length cut. The dashed line is the efficiency for keeping nuclear recoil events and the dotted line is the efficiency for keeping electron recoil events from \iso{39}{Ar} decay. The long dashed lines (horizontal) help to compare the efficiencies at the maximum figure of merit.}
    \label{fig:lengthCutFOM}
\end{figure}
The correlation between changing a cut on the length of the candidate event and the efficiency for simulated DM is shown in Fig.~\ref{fig:lengthCutFOM} for simulated DM and \iso{39}{Ar}. 
The simulation shows that cutting on the length of the event is a proxy for the F90 PID cut, a pulse shape discrimination parameter normally used in liquid argon detectors~\cite{Akimov:2019xdj,Akimov:2020pdx}, 
thus removing more energetic electromagnetic events. Setting a length cut of 48\,ns has a DM efficiency of about 80\% while reducing \iso{39}{Ar} background that pass the cut by about 60\%.   Using cleaner argon will greatly reduce the number of \iso{39}{Ar} events passing the data selection criteria due to the improved light output, without requiring any improvements to the analysis.  The impact of using cleaner argon is discussed further in Sec.\,\ref{sec:ccm200}. 

\subsection{Determine $\bm{T_0}$}
To search for coherent scattering from prompt $\pi^\pm$ decay and DM events that are assumed to come from $\pi^{0}$ decay, one of the most critical characteristics of the experimental setup is knowing when the first speed-of-light particle is expected to arrive at the detector, $T_0$.  
To determine $T_0$, a 1-liter EJ301 detector was placed inside of an adjacent flight path (FP3), to measure the gammas being produced from the protons hitting the target. Because of the high event rate, no particle identification could be made. However, with enough statistics the turn on can be measured, and knowing that the neutrons must first pass though a moderator, the turn on must be coming from gammas. This was verified by changing the amount of lead that was placed in front of the 1-liter EJ301 detector.
\begin{figure}[htp]
\centering
\includegraphics[width=0.48\textwidth]{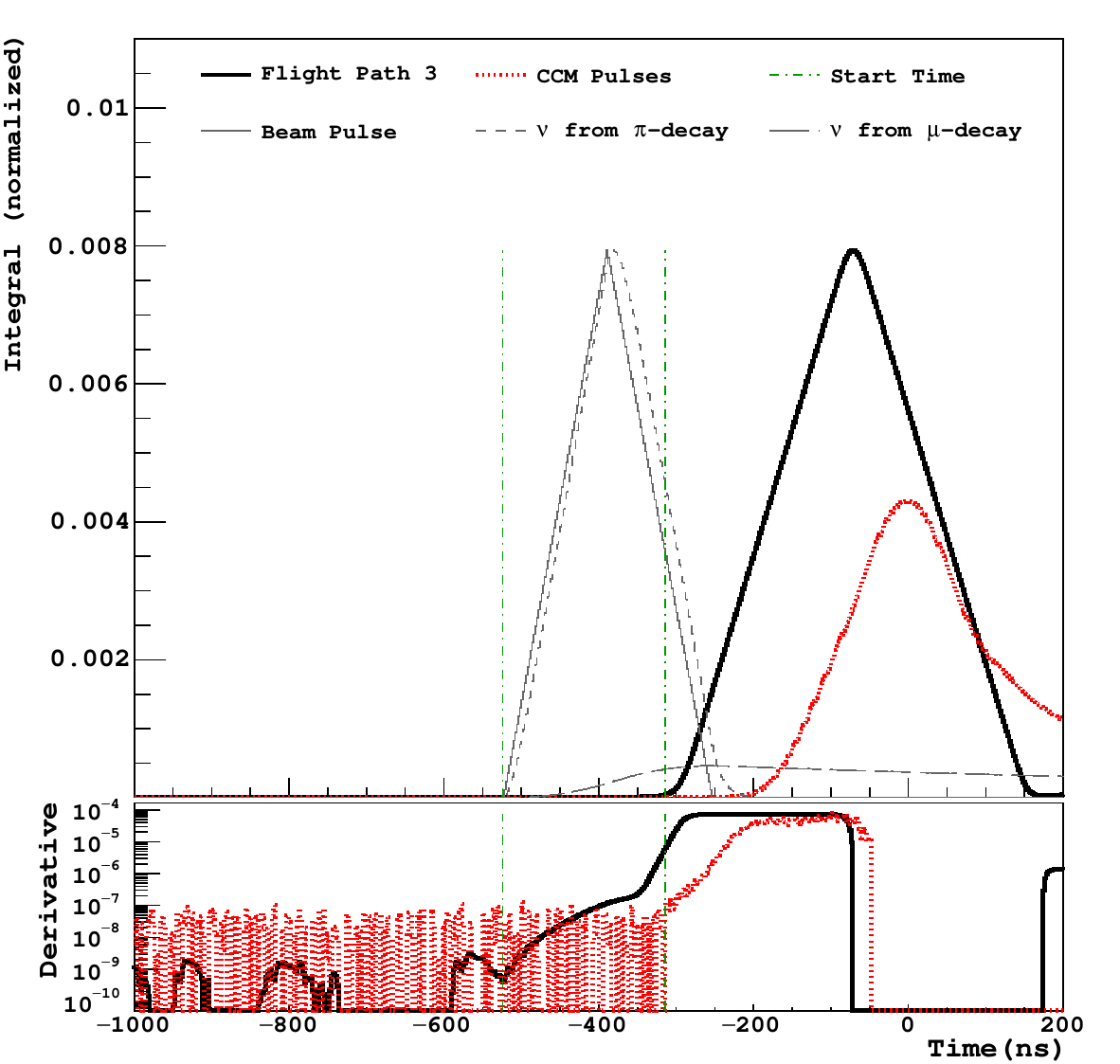}
\caption{Comparing the Flight Path 3 (FP3) and CCM event turn on to determine neutrino arrival time in CCM. (Top) The FP3 (solid thick), and CCM (dotted) accumulated waveforms as a function of time in the DAQ window. Both distributions are normalized to one. (Bottom) the derivative of the accumulated waveforms. The FP3 turn on is represented by the left most dot short-dashed line. The CCM turn on is represented by the right most dot short-dashed line. The thin solid, dashed, and long-dashed lines show the beam pulse, $\nu$ from $\pi$-decay, and $\nu$ from $\mu$-decay respectively, each starting at the FP3 turn on.  LDM from $\pi^0$ decay, with a quick lifetime of $8.5\times10^{-8}$\,ns, is represented by the beam pulse line.}
\label{fig:T0}
\end{figure}

 The results from measurements made to determine $T_0$ are shown in Fig.\,\ref{fig:T0}. The plot compares the accumulated waveforms from the FP3 EJ301 detector and the pulses observed in CCM after being corrected for cable length and detector distance difference. A first order derivative was used to determine when each detector first saw light. FP3 measured a time of -524\,ns and CCM measured a time of -314\,ns, a time difference of 210 ns.
 
Taking the triangular profile of the beam and assuming the width of the pulse is 280\,ns, about 90\% of the speed-of-light events produced from $\pi^0$ decays are reaching CCM before CCM starts observing a large turn on of beam related events. The beam region-of-interest (ROI), is defined in the Sec.\,\ref{sec:dmCuts} and is based on the measured efficiency.

\subsection{Selection Criteria for DM Search\label{sec:dmCuts}}
\begin{figure}[htp]
\centering
\subfloat[DAQ Time\label{fig:cut_progression_time}]{\includegraphics[width=0.4\textwidth]{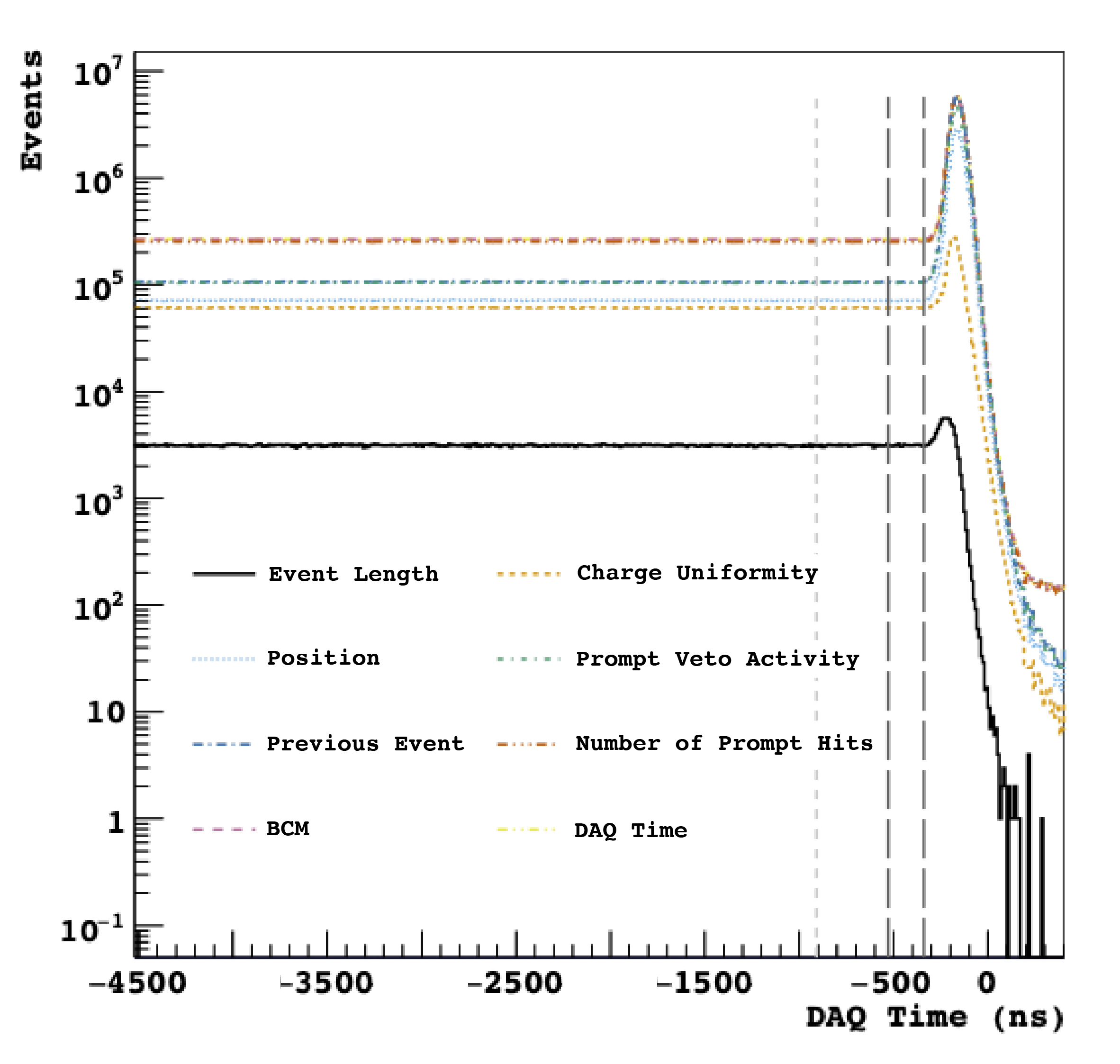}}\\
\subfloat[Reconstructed Energy\label{fig:cut_progression_energy}]{\includegraphics[width=0.4\textwidth]{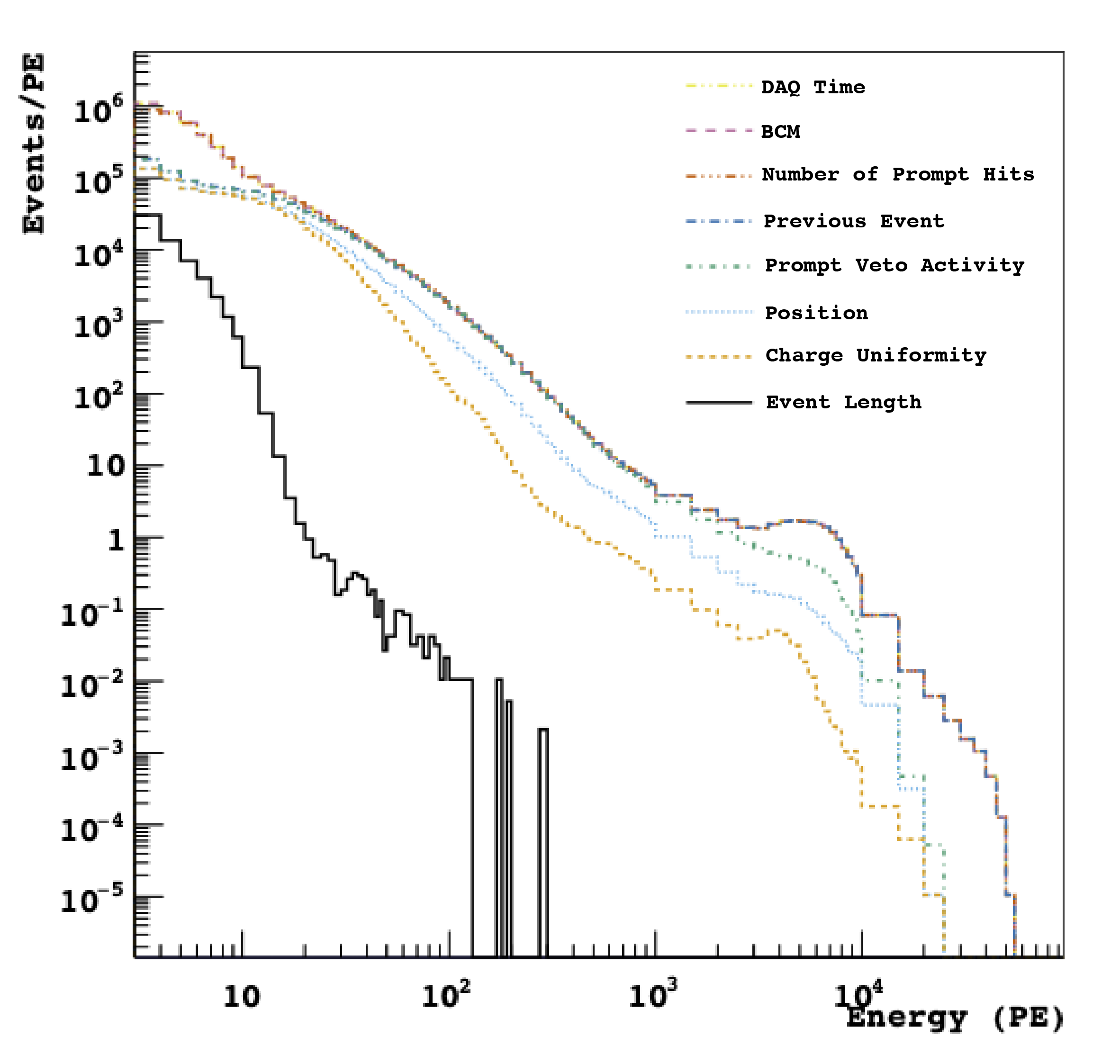}}
\caption{Progression of event selection criteria for both (a) DAQ time and (b) reconstructed energy. The region to the left of the short-dashed vertical line is the beam-out-of-time region and the region between the thick long-dashed vertical lines is the beam ROI. The reconstructed energy plot is for the beam-out-of-time region only.}
\label{fig:cut_progression}
\end{figure}

The following selection criteria are used to search for nuclear recoil events in the beam ROI defined as the DAQ time between -530 and -340\,ns, see Fig.~\ref{fig:cut_progression}.
A beam quality cut is also applied using the BCM signal, and this removes about 5\% of the beam triggers delivered. The signal arriving from the beam current monitor, used to create the trigger for the Lujan facility, is copied into the CCM DAQ. BCM slection criteria are based on the length, time, and integral of the beam current monitor waveform.

Triplet light, light originating from the decay of triplet excited states in $Ar_2^*$ (1.6$\mu$s), can potentially appear from a previous event.  A cut is applied by looking at previous candidate events that occurred in the same DAQ window. The ``previous event cut'' is energy dependent and is optimized to maximize sensitivity to \iso{57}{Co} over beam-off background.  The ``previous event cut'' also has its biggest effect on lower reconstructed energy (Fig.~\ref{fig:cut_progression_energy}), and is consistent with rejecting triplet light from a previous event.

To make sure the nuclear recoil occurs in the detector, a limit of no more than 3 veto hits in any 90\,ns window starting from 90\,ns before the start of the event to the end of the event or 90\,ns after the event (depending on which occurs first) is required.  The effect of the veto cut is shown in Fig.~\ref{fig:cut_progression_energy}. The veto cut is most prominent in the reconstructed energy region consistent with energy deposit from cosmic muons.

A fiducial volume with a radius of 95\,cm and a height of $\pm$40\,cm is used. The position of the event is determined by a simple weighted average of the observed PMTs where a PMT's weight is the amount of charge squared observed in the first 20\,ns of the event. The energy dependence on the position cut is shown in Fig.~\ref{fig:cut_progression_energy} and is consistent with higher PE/keV the closer the event occurs near the edge of the detector.

To insure an event is treated equally between PMTs, and thus removing events that come from outside the detector that do not interact in the veto region, a charge uniformity cut is applied. The charge uniformity $\mathcal{U}$ is defined as, $\mathcal{U} = E_{\text{largest PMT}} / \bar{E}_{\text{all PMT}}$,
where $E_{\text{largest PMT}}$ is the largest PE observed in the PMTs that measured charge in the prompt region of the event, and $\bar{E}_{\text{all PMT}}$ is the average charge observed across all visible PMTs in the prompt region.
A maximum value of $\mathcal{U} = 2.5$ is used. The value is determined by looking at the efficiency of \iso{57}{Co} and \iso{22}{Na} data. The \iso{22}{Na} and \iso{57}{Co} excesses over background increase by about 5\% with the charge uniformity cut.

Lastly, as stated in Sec.~\ref{sec:simDMResponse}, the event length can be used as a proxy for particle identification. DM and \iso{39}{Ar} simulations were used to create a figure of merit (FOM) that is shown in Fig.\,\ref{fig:lengthCutFOM}. The FOM was calculated by taking the ratio of the DM efficiency divided by the square root of the \iso{39}{Ar} efficiency. A cut of 48\,ns was chosen and, as Fig.~\ref{fig:cut_progression_energy} shows, the event length cut has its biggest effect in the mid-to-high energy reconstruction.

 The efficiencies for each selection criteria are shown as a function of DAQ time in Fig.\,\ref{fig:cut_progression_time}. All the selection criteria have a flat efficiency in the beam-out-of-time region, left of the short dashed vertical line. The charge uniformity and event length selection criteria have the largest effect (i.e., rejection) on the events just to the right of the beam ROI.

\begin{figure}[htp]
    \centering
    \includegraphics[width=0.48\textwidth]{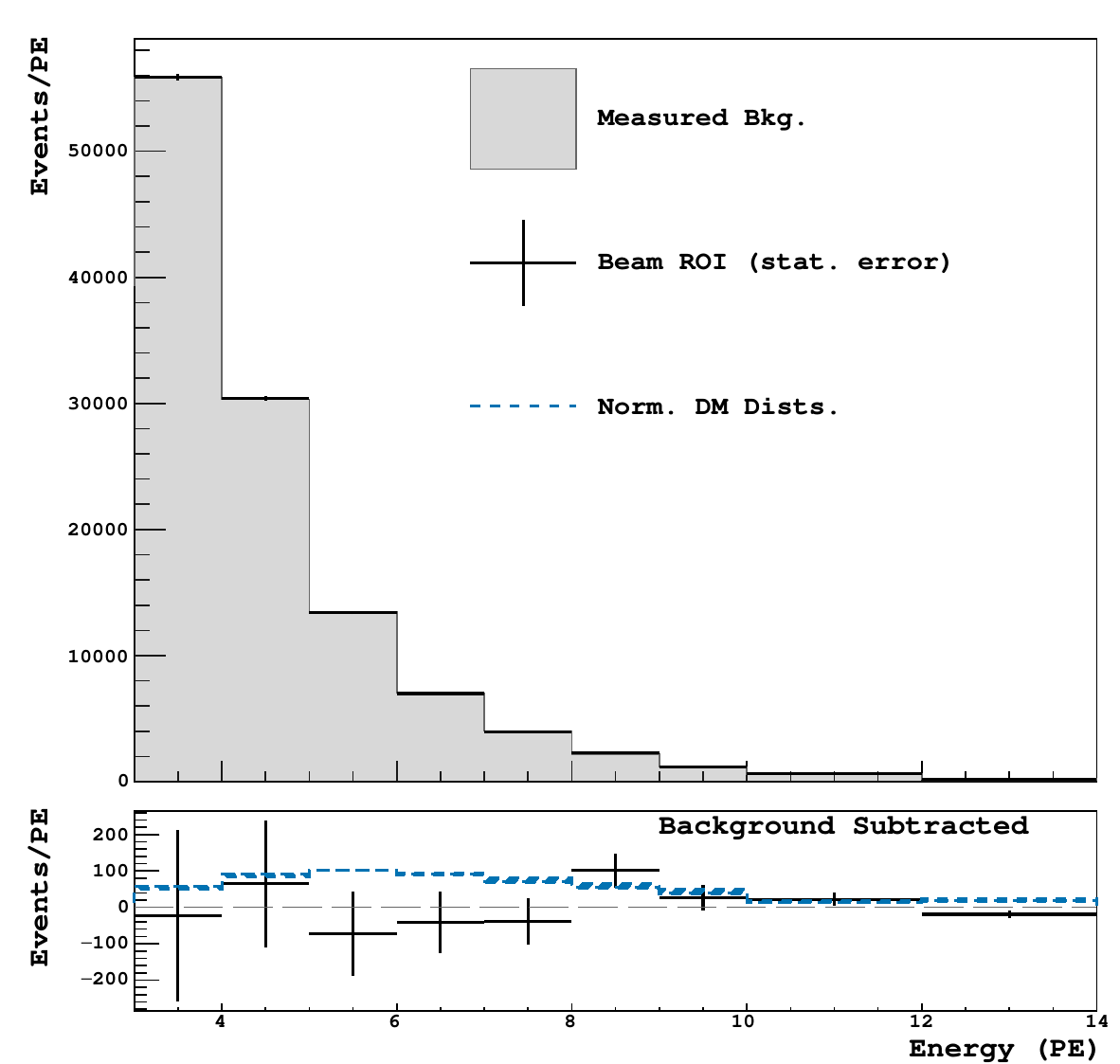}
    \caption{The top frame shows the reconstructed energy distribution after all selection criteria are applied. The background prediction based off the beam-out-of-time window is the shaded region, and measured data in the beam signal region of interest are the solid lines. 
    The bottom frame shows the background subtracted distribution along with a blue line that is arbitrarily normalized to show the shape of the expected DM distribution.  The thickness of the blue line shows the variation due to 383 different $m_{_V}$, $m_{\chi}$ mass combinations.}
    \label{fig:dataDist}
\end{figure}

The predicted background and beam ROI data observed for $17.894\times10^{20}$\, POT, or  56,860,679 triggers, is shown in Fig.~\ref{fig:dataDist}.
The background was calculated by averaging the data observed in the beam-out-of-time region and comparing it to the data occurring in the beam ROI time window. The beam-out-of-time window is 22 times bigger than the 190\,ns beam ROI. The beam-out-of-time window was split into 190\,ns time bins and no variation above statistical fluctuations was observed.

The background prediction is consistent with observations made in the beam ROI.  The observed data is 115,005 events with a net after subtraction of $16.5\pm338.4$ events, where the uncertainty (338.4) is the statistical error on the background. A comparison is made in the bottom frame of Fig.\,\ref{fig:dataDist} between the background subtracted distribution and the expected shape of an arbitrarily normalized DM distribution (thick blue line).  The thickness of the blue line shows the variation due to 383 different $m_{_V}$, $m_{\chi}$ mass combinations.

\subsection{Systematic Uncertainties\label{sec:sys}}
As stated in Sec.\,\ref{sec:dmCuts} the background is computed from the beam-out-of-time region and a background systematic uncertainty is not observed to be above the statistical fluctuations. Therefore, the only systematic errors included in the DM fit are those on the DM prediction itself that is not DM model related.

There are three systematic errors that are included in the analysis: (i) the uncertainty in the amount of POT which was measured to an accuracy of 0.7\%, (ii) the uncertainty coming from the quenching factor of nuclear recoil events in liquid argon, and (iii) propagating the covariance matrix from the optical model parameters.
As mentioned in Sec.~\ref{sec:radioactiveSources}, the quenching factor used in this analysis is $0.272\pm0.035$ and corresponds to a fractional uncertainty of 12.0\% in reconstructed DM events.  

To propagate the optical model covariance, the covariance matrix was probed 450 times to simulate DM events with optical model properties that are consistent with the covariance matrix. Each ``throw'' is put through the selection criteria to create the optical model event covariance matrix around the central value simulation. The optical model fractional error is 18.8\% in reconstructed DM events.

The total fractional error is 22.6\% in reconstructed DM events. This compares with the 5\% statistical fractional error on the predicted background. Although the fractional error on the reconstructed DM events is about four times higher than the statistical fractional error, the statistical errors dominate the fit because one would need about 400 DM events before the total systematic error is about the same as the statistical error, and the background subtraction is only 16.5 events. Even so, the systematic uncertainties are included in generating the sensitivity and confidence limit discussed in Sec.~\ref{sec:cl}.

\section{Confidence Limits on LDM\label{sec:cl}}

\begin{figure}[htbp]
    \centering
    \includegraphics[width=0.48\textwidth]{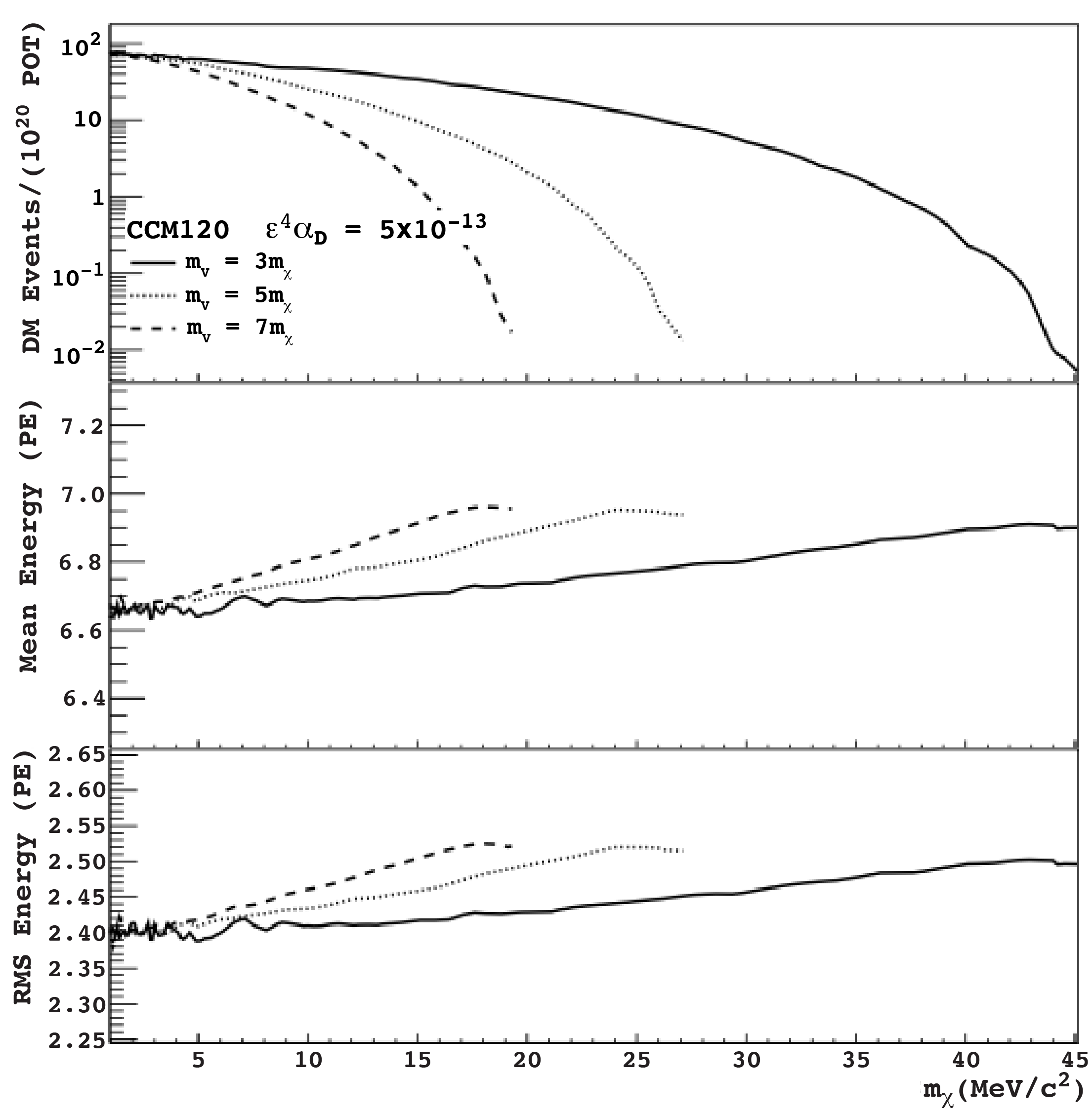}
    \caption{For the vector portal model, (top) the number of predicted DM events for $m_V = 3m_\chi$, $m_V = 5m_\chi$, and $m_V = 7m_\chi$ represented by solid, dashed, and dotted lines respectively. The predicted mean (middle) and RMS (bottom) PE distributions are displayed for each of the $m_V/m_\chi$ ratios shown on top. Each plot was made with $\varepsilon^4\alpha_{_D} = 5\times10^{-13}$, and after all selection criteria were applied.}
    \label{fig:dmDist}
\end{figure}

The mass parameter space for $0.3 < m_V < 136$\,MeV and $0.1 < m_\chi < 68$\,MeV was scanned to generate 2D confidence limits.  The predicted number of events and reconstructed distributions change across the mass parameter space for the vector portal model, and are shown in Fig.\,\ref{fig:dmDist}.

The overall effect is that the higher the ratio of $m_V/m_\chi$, the number of predicted events decreases, the mean reconstructed energies increase, and the root mean squares of the reconstructed energy increase.  A comparison between different $m_V/m_\chi$ ratios in the parameter $Y$ from the results of the confidence level test is also shown in  Fig.~\ref{fig:dmDist}. Higher $m_V/m_\chi$ ratios result in a lower $Y$ value. The same offsets between different $m_V/m_\chi$ values are observed in MiniBooNE full nucleon similar to what is seen in CCM120~\cite{Aguilar-Arevalo:2018wea}.

\begin{figure*}[htp]
    \centering
    \subfloat[\label{fig:dmCL_vector}vector portal]{\centering \includegraphics[width=0.48\textwidth]{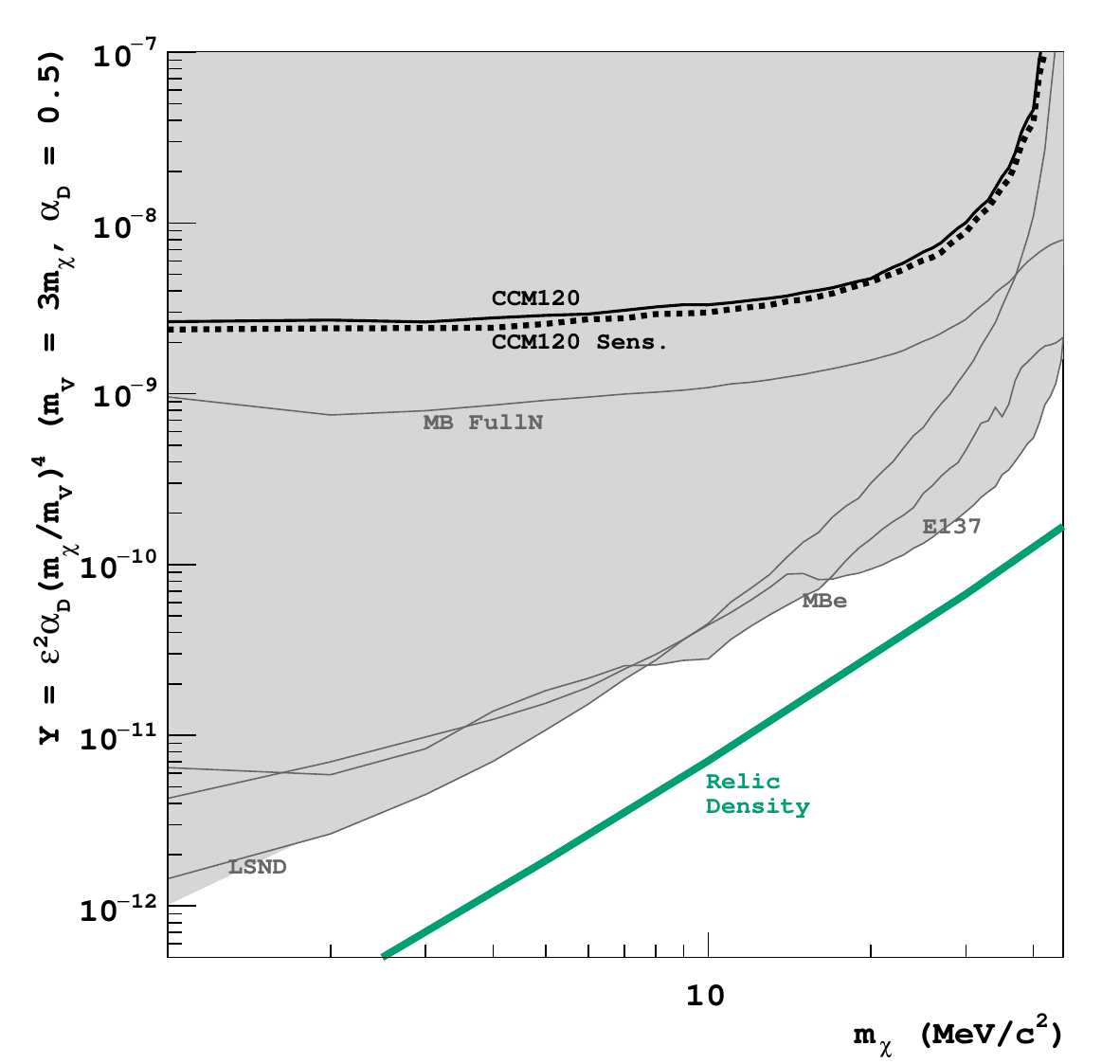}}
    \subfloat[\label{fig:dmCL_leptophobic}leptophobic]{\centering \includegraphics[width=0.48\textwidth]{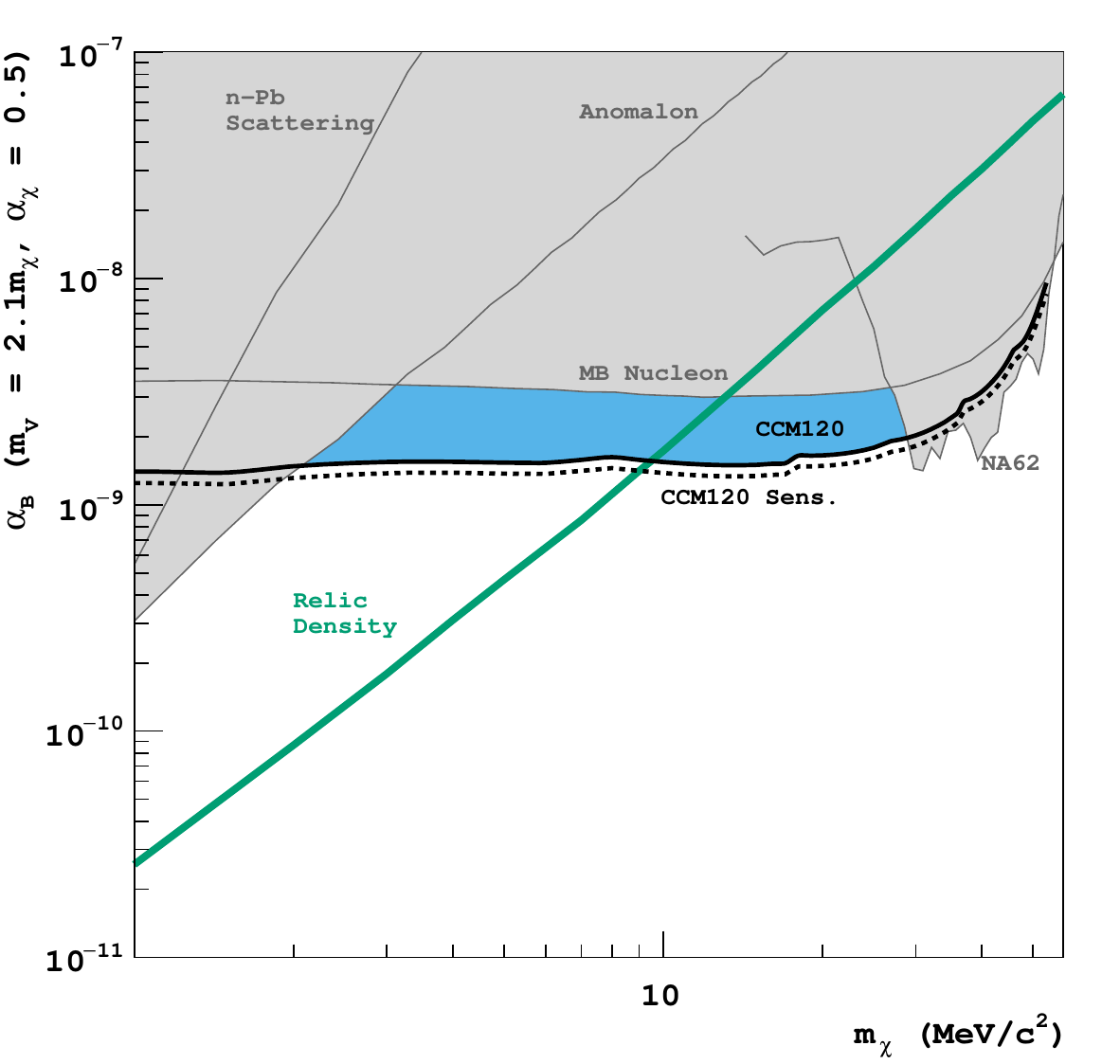}}
    \caption{The median sensitivity (dashed) and 90\% confidence limit (solid) results from CCM120 compared to current limits. (a) Vector portal model with limits from previous neutrino experiments, LSND, E137, and MiniBooNE, are displayed to show current limits from other experiments that are sensitive to $\varepsilon^4\alpha_{_D}$.  See Refs.~\cite{Batell:2014mga,Banerjee:2017hhz,Essig:2017kqs,deNiverville:2016rqh,Lees:2014xha,Aguilar-Arevalo:2018wea,Aguilar-Arevalo:2017mqx,Marsicano:2018glj,NA64:2019imj} for an explanation of the previous limits. (b)~leptophobic model. See Refs.~\cite{Aguilar-Arevalo:2018wea,Aguilar-Arevalo:2017mqx,Batell:2014yra,Coloma:2015pih,Dror:2017ehi,Berlin:2018bsc,Barbieri:1975xy} for an explanation of the previous limits.}
    \label{fig:dmCL}
\end{figure*}
The confidence level (CL) is calculated using the same frequentist method used for the MiniBooNE DM and oscillation analyses~\cite{Aguilar-Arevalo:2017mqx,Aguilar-Arevalo:2018wea,Aguilar-Arevalo:2020nvw}. The CL test uses the Feldman-Cousins approach, where the test statistic is a delta log-likelihood where the log-likelihood function is Gaussian. The uncertainties are incorporated assuming Gaussian symmetric errors. They are added to the statistics of the measured background in quadrature. In Fig.\,\ref{fig:dmCL}, the 90\%\,CL is the black solid line labeled CCM120, while the median sensitivity to background only is the black dashed line labeled CCM120 Sens.  The results of the 90\%\,CL are shown in Fig.\,\ref{fig:dmCL} for both the vector portal and the leptophobic models.

The vector portal assumes $m_V/m_\chi = 3$ and $\alpha_{_D} = 0.5$, while the leptophobic model assumes $m_V/m_\chi = 2.1$ with $\alpha_\chi = 0.5$. The vector portal slice is chosen to be consistent with previous publications. For the leptophobic model, a smaller mass ratio is chosen to show a part of the parameter space for which the complex scalar relic density line is not yet ruled out by existing limits. As $m_V/m_\chi$ increases the relic density line moves to larger required coupling compared to the CCM limit. Decreasing the value of $\alpha_{_D,\chi}$ reduces the strength of CCM relative to limits that do not rely on scattering signals, while also moving the relic density line to larger couplings $\alpha_{_B}$ and $\epsilon$.

 The current exclusion space for the vector portal model~\cite{Batell:2014mga,Banerjee:2017hhz,Essig:2017kqs,deNiverville:2016rqh,Lees:2014xha,Aguilar-Arevalo:2018wea,Aguilar-Arevalo:2017mqx} is shown in Fig.\,\ref{fig:dmCL_vector} and highlights the experiments that are sensitive to $\varepsilon^4\alpha_{_D}$, (LSND, E137, and MiniBooNE). MiniBooNE has two limits: (i)~looking for DM having quasi-elastic and inelastic scatters off nucleons (MB FullN) and (ii)~elastically scattering off electrons (MBe). CCM120 is the second of these types of searches to be sensitive to DM that does not require interactions with electrons; MB FullN is the other one.
 The current exclusion space for the leptophobic model~\cite{Batell:2014yra,Coloma:2015pih,Dror:2017ehi,Berlin:2018bsc} is shown in Fig.\,\ref{fig:dmCL_leptophobic} and shows the most conservative and least model dependent anomalon limit~\cite{Dror:2017ehi}.
 
With about 1.5 months of data and contaminated liquid argon that reduced the 128\,nm attenuation length to about 50\,cm, the CCM120 line is close to the MB FullN limit for the vector portal model.  At the same time, CCM120 excludes new parameter space in the leptophobic model where $\sim2.0 < m_\chi < 30$\,MeV. The CCM120 limit also covers the complex scalar relic density in the leptophobic model above $m_\chi \sim$9\, MeV.

\section{Future Upgrades \label{sec:ccm200}}


\begin{figure}[tbh]
   \centering
   \includegraphics[width=3.5in]{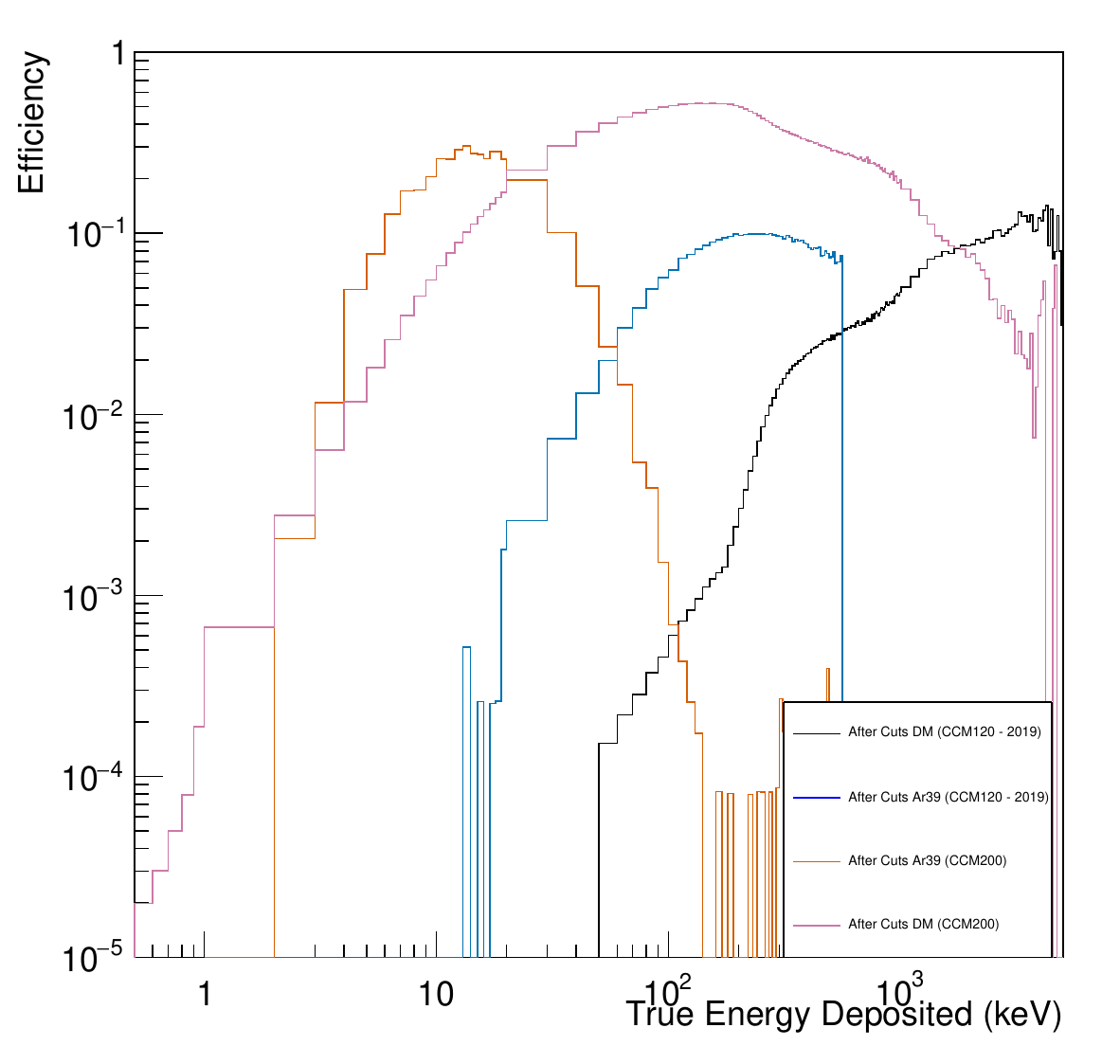} 
   \caption{CCM200 expected improvements in efficiency from simulations. The biggest effect is from the improved attenuation length from the clean argon.  The pink (DM) and orange ($^{39}$Ar) lines are for clean/filtered LAr expected with CCM200, while the black (DM) and blue ($^{39}$Ar) lines are with the contaminated LAr in CCM120. When convoluting this with the expected spectrum, the CCM200 energy integrated DM efficiency is about 25\% while only 1\% for $^{39}$Ar decay.  }
\label{fig:cleanAr}
\end{figure}

\begin{figure*}[htp]
    \subfloat[\label{fig:dmReach_vector}vector portal]{\centering
    \includegraphics[width=0.48\textwidth]{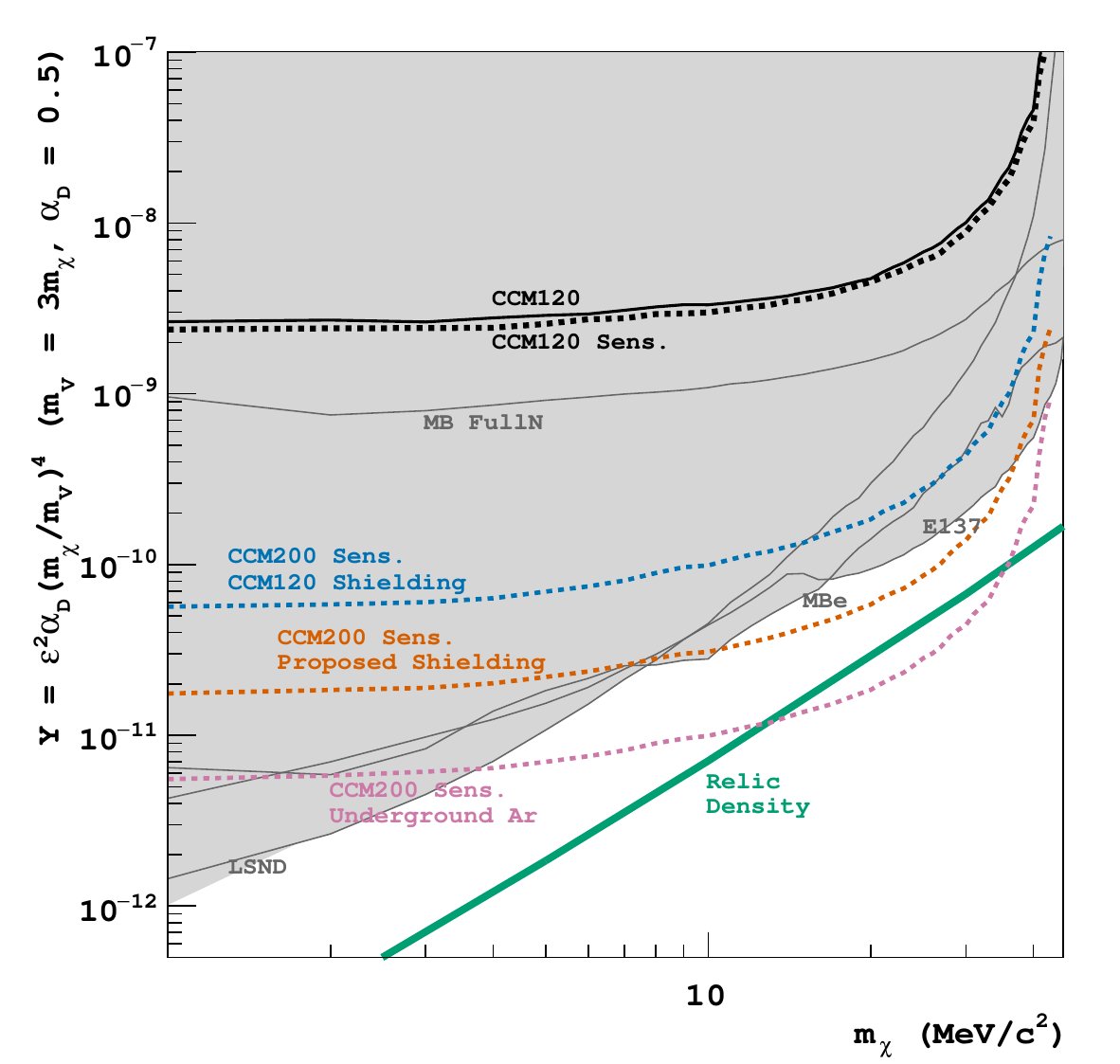}}
    \subfloat[\label{fig:dmReach_leptophobic}leptophobic]{\centering
    \includegraphics[width=0.46\textwidth]{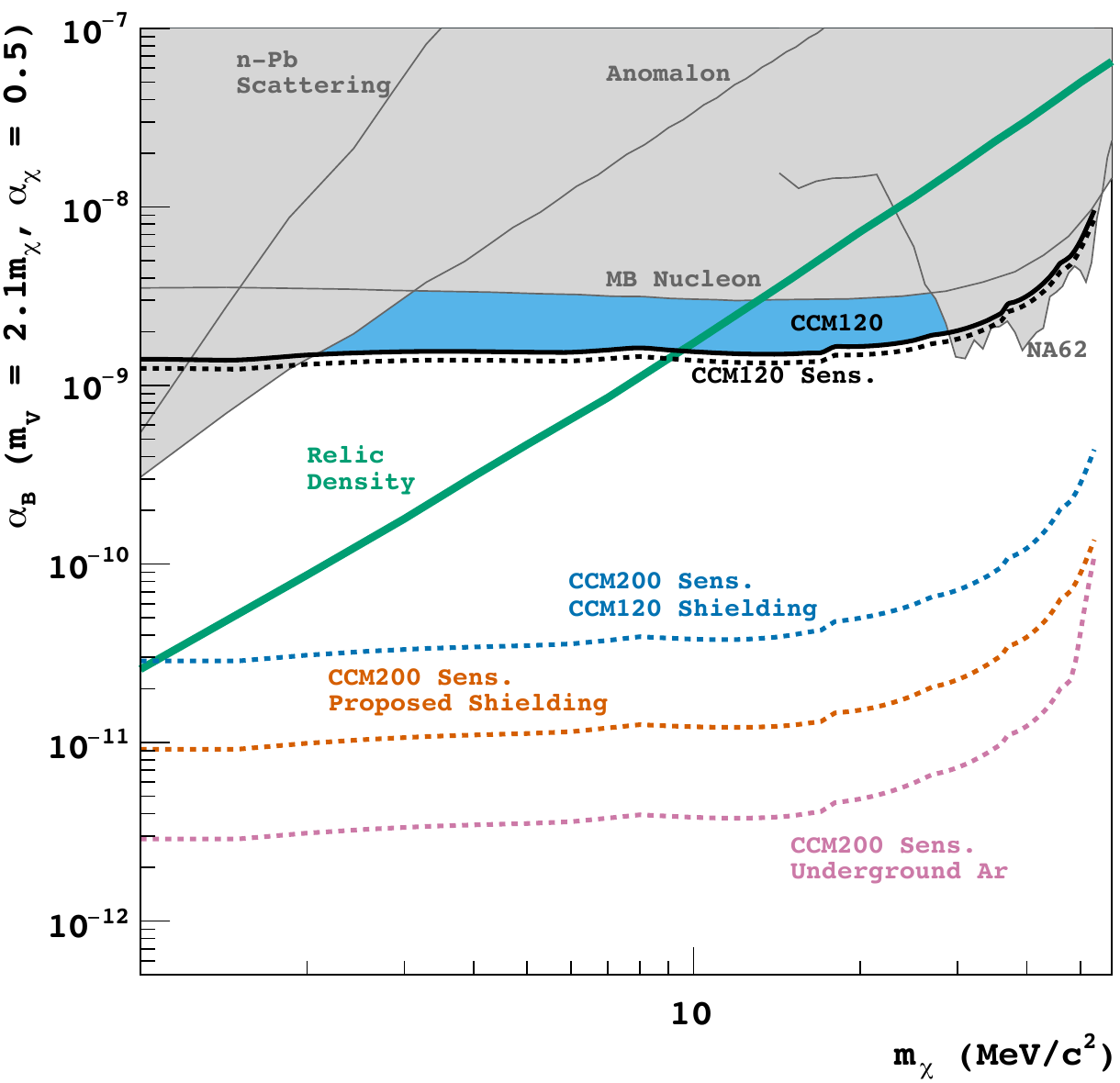}}
     \caption{The blue line is the CCM200 DM full fit sensitivity with 10 keV threshold for a three year run ($2.25\times10^{22}$~POT) and is based on the most recent CCM200 simulation with clean LAr and improved instrumentation.  The background model is taken from the measured CCM120 rates and shape that are adjusted for the effects of the assumed improvements.  The orange CCM200 sensitivity line includes the effects of upgraded shielding. The pink line could be achieved by using isotopically pure LAr to further reduce the $^{39}$Ar backgrounds, or significant background rejection with improved analysis techniques. Limits from previous neutrino experiments are shown in Fig.\,\ref{fig:dmCL}.}
    \label{fig:dmReach}
\end{figure*}

\begin{figure}[htp]
    \centering
    \includegraphics[width=0.48\textwidth]{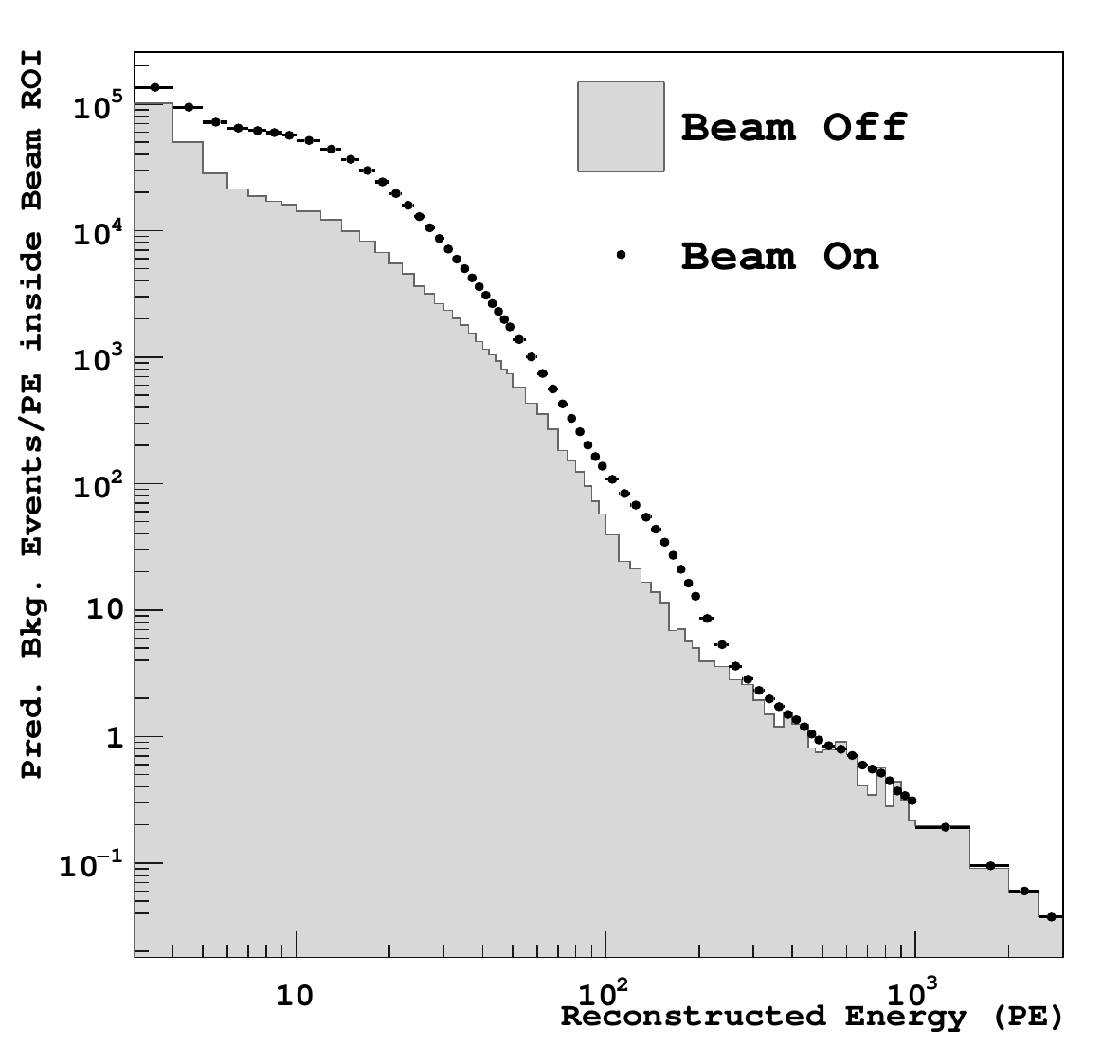}
    \caption{Comparing the beam-out-of-time DAQ window rates from when the beam is on to when the beam is off. The beam on rate is about 3.5 times higher than the beam off rate. }
    \label{fig:beamOnBeamOffRate}
\end{figure}

A significant enhancement in DM sensitivity will be achieved with the new CCM200 detector that will begin running in 2021.  It will have 200  8-inch PMTs, twice as many veto PMTs for better external background rejection, highly efficient evaporative coated foils, and filtered and recirculated LAr.  The biggest impact will come from the filtered LAr, with an expected factor of 1000 increase in DM efficiency at the lowest energies.  Figure~\ref{fig:cleanAr} shows the expected efficiency enhancement for CCM200 from GEANT4 detector simulations where the LAr is assumed to have no significant O$_2$ or H$_2$O contamination (clean).  The biggest effect is at low energy where a goal of 10 keV threshold will significantly improve the DM search sensitivity. With the lower threshold, prompt neutrinos from $\pi^+$ decays will have to be subtracted. Just as important as the increased signal efficiency, is the expected decrease in $^{39}$Ar events, as the efficiency for these event decreases when applying the same data selection criteria.  This is due to the increase in light output, making identification of triplet (delayed) light more efficient.  The $^{39}$Ar efficiency for CCM120 goes from 26\% to only 1\% for CCM200.  This represents a significant reduction in background from this source, with the expectation of about 3000 events a year for CCM200 after all selection criteria are applied.  The \iso{39}{Ar} efficiency could decrease further with improvements in analysis, leveraging machine learning (ML) and other advanced techniques.  In fact, CCM200 is ideal for ML techniques as the algorithms will be trained (unsupervised) on the backgrounds measured with out-of-time beam data and do not rely on simulated backgrounds that are only approximations and usually limit such techniques.

The projected CCM200 vector portal model and leptophobic DM sensitivities are shown in Fig.\,\ref{fig:dmReach} where the blue line shows the expected improvements for CCM200 assuming the current shielding, and assuming a nominal three year run collecting $2.25\times10^{22}$~POT.  If this significant improvement is achieved, then the next step to further improving the DM sensitivity will be to reduce backgrounds with additional shielding. The first shielding upgrade will add more than 160 tons of steel in the upstream areas close to the target.  This will further slow down neutrons and increase the prompt signal window region free of such backgrounds by as much as $\sim$200 nsec.  The second shielding upgrade will enclose CCM200 detector with 101 tons of steel with roof and side shielding.  The CCM120 beam-on and beam-off rates for out of time events is shown in Fig.\,\ref{fig:beamOnBeamOffRate}.  The excess is due to $\gamma$-rays from neutron activation in nearby structures (see~\ref{app:gamma-rays}).  With the steel enclosure shielding, it is predicted that the 3.5 beam-on excess will be reduced to the point where $^{39}Ar$ becomes the dominant out-of-time background.  In this scenario, the limiting background is estimated to be  $9120 \pm 96$ in three years, assuming a 1\% contamination rate based on the CCM200 simulation efficiency shown in Fig.\,\ref{fig:cleanAr}. This represents the floor intrinsic background rate.  The DM sensitivities for this are shown in Fig.\,\ref{fig:dmReach}, with the orange sensitivity line showing about two orders of magnitude increase in sensitivity over CCM120 limits, and a factor of three improvement over CCM200 without the shielding.  This begins to sample untested parameter space around a DM mass of 20\,MeV. At this level, we begin to reach beyond the sensitivity of other experiments, such as LSND, E137, and MiniBooNE.

To probe the relic density limits and beyond, an extremely low background rate of  $100 \pm 10$ events will be required for a full three year run, and is shown in Fig.~\ref{fig:dmReach} as the pink line. If a nearly background free region can be attained by using both timing and energy selection criteria, then such a limit should be achievable. This could be possible with isotopically pure argon from underground sources, PSR reduction in beam width timing, and other analysis improvements. This is one of the goals of the 2021 beam run and will be explored over the coming years with further research.

The complex scalar relic density in the leptophobic model below 60\,MeV will be completely covered for all invisible decay parameter combinations with CCM200 by just improving the quality of the liquid argon from what CCM120 had. Improving the shielding and the low background argon will cover relic density lines from other types of DM particles.

\section{Conclusions}
The CCM120 experiment proved successful on many fronts.  It tested new technology to be used in CCM200 and other neutrino experiments, and it explored new regions of the LDM parameter space, most notably in the leptophobic model.   The shakedown of the CCM120 detector during the 3-month engineering run revealed the need for improvement in two broad areas:  (1) the beamline and shielding, and (2) the detector and offline analysis.  

As in most engineering runs, the analysis was stretched to its limits.  Some of the improvements were made during the run and many improvements (e.g., software algorithms) occured after the run was finished.  Early on, the analysis showed that there was an early window of opportunity of $\sim$190\,ns where neutrinos and DM can be observed before the wave of beam neutrons arrived.  This signal region sensitivity was limited by out of time backgrounds from neutron-induced $\gamma$ rays and $\iso{39}{Ar}$ decays.  Both of these issues are being addressed with upgraded detector shielding in the upcoming CCM200 run (2021).  

There are two possible accelerator-related improvements that could extend the physics reach for CCM200.  The first is narrowing the bunch time baseline from 250\,ns to  $\sim$100\,ns.  The second improvement would be additional steel shielding near the target to attenuate and delay the neutrons on their way to the CCM detector.  Addressing both these issues will improve the signal-to-noise ratio for identifying candidate LDM and $\nu_s$'s in the beam-related background free signal window.

On the detector side, a number of improvements have been made, including more fiducial-volume and veto PMTs, filtering and recirculating the LAr, and using evaporated, vs. painted, TPB coating on the foils. Addressing these issues will improve the light output, efficiency, and energy resolution.

CCM120’s biggest success is its ability to use an engineering run with 1.5 months of analyzable data to set mass limits on LDM using the vector portal and leptophobic models.  In particular, new limits for LDM in the mass range $9<m_{\chi}<50$\,MeV in the leptophobic model are shown in Fig.\,\ref{fig:dmCL}.  With the improvements already implemented in CCM200 (filtered LAr, more shielding, etc.), it is reasonable to expect that new mass limits for LDM will be forthcoming using the vector portal and leptophobic models.   

\section{Acknowledgements}

We acknowledge the support of the Department of Energy, the National Science Foundation, Los Alamos National Laboratory LDRD funding, and support from PAPIIT-UNAM grant No. IT100420.  We also wish to  acknowledge the support of LANSCE Lujan center and Accelerator Operations and Technology (AOT) division.  This research used resources provided by the Los Alamos National Laboratory Institutional Computing Program, which is supported by the U.S. Department of Energy National Nuclear Security Administration under Contract No.\,89233218CNA000001. Special thanks to Deion Fellers for setting up the germanium detector and collecting the gamma ray data near the CCM120 detector.


\section{Appendices}
\subsection{Gamma-Ray Measurements near the CCM120 Detector,\label{app:gamma-rays}}
\begin{figure} 
  \centering
  \includegraphics[width=0.48\textwidth]{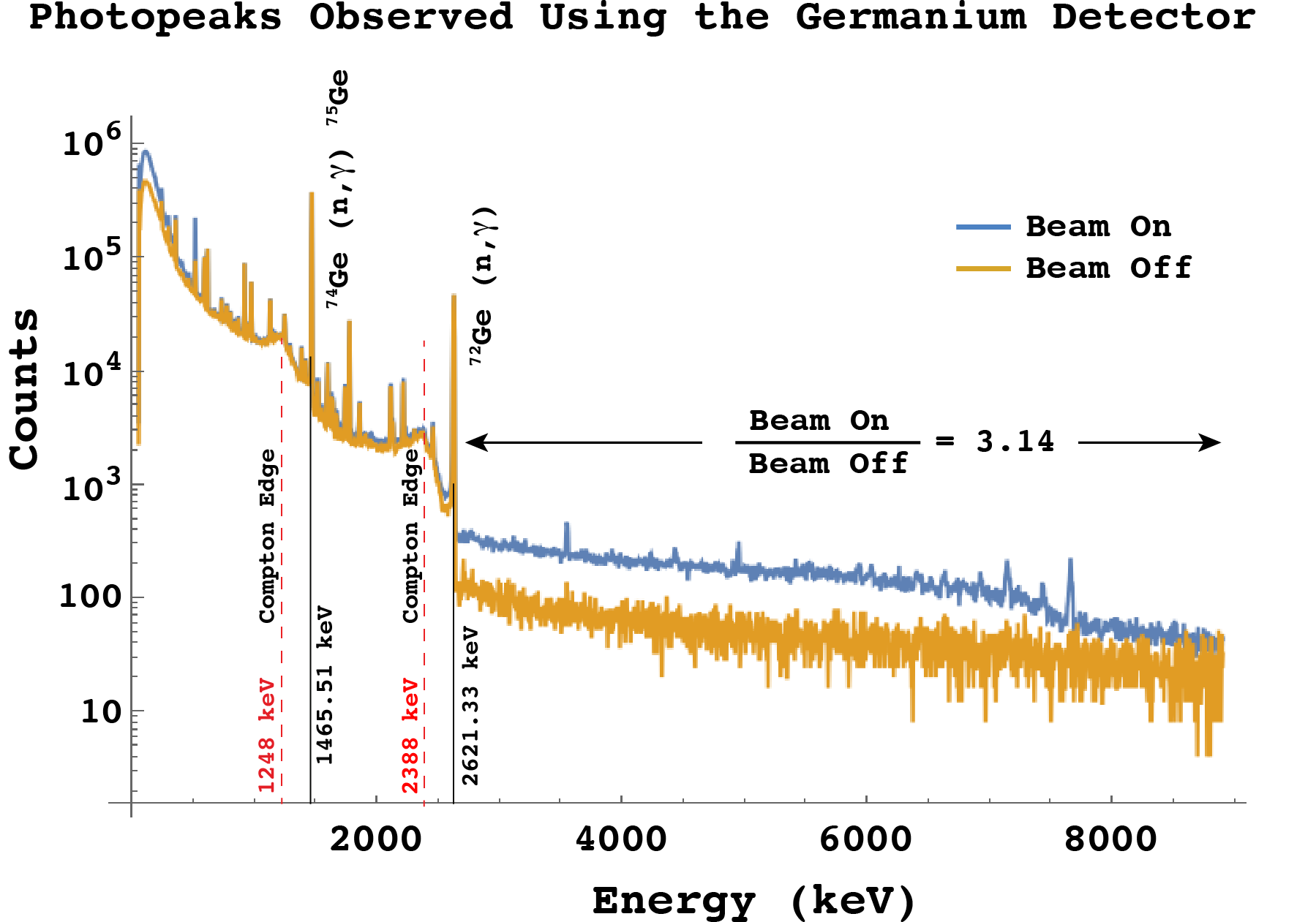}
  \caption{The gamma ray spectra for both beam-on and beam-off conditions were recorded near the CCM120 detector.  The spectra were normalized to 17.25 hours of live time so comparisons could be made.  Approximately 19 photopeaks were observed between 50-2700 keV with the same rate for beam-on and beam-off.   However, above 2700~keV, the beam-on to beam-off ratio was $\sim$3:1, consistent with the observations found in Fig.~\ref{fig:beamOnBeamOffRate}.}
  \label{fig:BeamOnBeamOffNormalized}
\end{figure}

Measurements of the $\gamma$-ray backgrounds were recorded immediately following the engineering run.  A germanium detector was set up near the CCM120 detector facing the tungsten target, and data were recorded for both beam-on and beam-off conditions similar to the engineering run.  The beam-on data had a live time of 17.25 hours and the beam-off data were normalized to that ($\times 4.01$) so comparisons could be made between the two spectra (see Fig.~\ref{fig:BeamOnBeamOffNormalized}).  

Approximately 19 photopeaks were observed from 50-2700 keV with prominent peaks (peaks 12 and 19) located at 1465 and 2621 keV respectively.  These were determined to be neutron-induced photopeaks occurring internally in the germanium detector. Their respective Compton edges (i.e., highest energy $e^-$ recoils) are easily identified in the plot and located at the expected energies.
The 19 photopeaks from 50 - 2700 keV appear to have the same rates between beam-on and beam-off except for peak 4.  Peak 4 appears to have 4 times the rate during beam-on conditions.  Upon closer investigation, it is believed that peak 4 is really two peaks and the beam-on peak has a shorter lifetime than the beam-off peak near by.  Another interesting feature is observed in the high energy region from 2700-9000 keV where no photopeaks are observed and the ratio of beam-on to beam-off $\gamma$-ray backgrounds is $\sim$ 3:1.  This feature is consistent with the observations shown in Fig.~\ref{fig:beamOnBeamOffRate} where a ratio of 3.5 is observed.


\subsection{Coherent DM Scattering}

Coherent vector portal DM scattering off some nucleus $A$ with a number of protons $Z$ can be written as \cite{deNiverville:2015mwa}
\begin{equation}
    \frac{d\sigma_{\chi _A}}{d T} = \frac{2\pi Q_A^2\alpha_{_D} \alpha_{_{\mathrm{EM}}}\epsilon^2 m_{_A} (T - 2E_\chi)^2}{(E_\chi^2-m_\chi^2)(q^2+m_V^2)^2},
    \label{eq:c_dm_scatter}
\end{equation}
where $A$ is assumed to be a scalar particle, $m_A$ is the mass of the nucleus, $T$ is the recoil energy, $E_\chi$ is the incoming DM particle's energy, and $q^2 = 2 m_A T$ is the momentum exchange. The effective charge of the nucleus is given by
\begin{equation}
    Q_A = F_\mathrm{Helm}(q^2) Z,
    \label{QA}
\end{equation}
where $F_\mathrm{Helm}$ is the Helm's form factor \cite{Helm:1956zz,Duda:2006uk}. Equation~\ref{eq:c_dm_scatter} makes corrections to a previous treatment of coherent scattering in the literature found in \cite{deNiverville:2015mwa} and is consistent up to spin-dependent effects with \cite{Dutta:2019nbn}.

The coherent cross section in the leptophobic case is very similar to Eq.~\ref{eq:c_dm_scatter}, only differing in overall coupling factors \cite{deNiverville:2015mwa},
\begin{equation}
    \frac{d\sigma_{\chi _A}}{d T} = \frac{2\pi Q_B^2 \alpha_{_B} \alpha_{_D} m_{_A} (T - 2E_\chi)^2}{(E_\chi^2-m_\chi^2)(q^2+m_V^2)^2},
    \label{eq:cl_dm_scatter}
\end{equation}
where $\alpha_{_B} = \frac{g_B^2}{4\pi}$ and 
\begin{equation}
    Q_B = F_\mathrm{Helm}(q^2) A.
    \label{eq:QB}
\end{equation}
Note that the $V_B$ couples to baryon number rather than hypercharge, so the number of protons $Z$ is replaced by the total number of nucleons (baryons), $A$.

The coherent scattering cross sections in Eq.\,\ref{eq:c_dm_scatter} (vector portal model) and Eq.\,\ref{eq:cl_dm_scatter} (leptophobic model) determine the shape of the sensitivity curves shown in Figs.\,\ref{fig:dmReach} (a,b). The factor of $(q^2+m_V^2)^2$ in the denominator divides the cross section into regimes that are independent of the dark photon mass and scale as $q^{-4}$ when $q^2\ll m_V^2$, and regimes that scale as $m_V^{-4}$ when the opposite condition is true. This is reflected in the plots as a flattening of the sensitivity curve at low $m_{_V}=3m_\chi$ because the cross section becomes independent of $m_{_V}$. The exact point where this change-over occurs depends on the minimum value of $q^2$, and therefore the recoil energy threshold adopted. At large values of $m_{_V}$, the curves rapidly become dominated by suppression in the production due to $m_{_V} \approx m_{\pi^0}$ (see Eq.~\ref{eq:mesondecay}), and the sensitivity to the model weakens.

\subsection{Additional Physics Goals: Inelastic neutrino-argon scattering} 
Neutrinos produced at the Lujan Center can also scatter off the argon nucleus in the CCM detector via CC or NC inelastic scattering process that are relevant for supernova neutrino physics since the decay-at-rest neutrino spectrum produced at the Lujan Center significantly overlaps with the expected supernova neutrino spectrum. Incidentally, the detection of the burst of tens-of-MeV neutrinos from the galactic core-collapse supernova is one of the primary physics goals of the DUNE experiment~\cite{Abi:2020lpk}. But the inelastic neutrino-nucleus cross sections in this tens-of-MeV regime are quite poorly understood. There are very few existing measurements, none at better than the 10\% uncertainty level, and no measurement on the argon nucleus is performed to date. As a results, the uncertainties on the theoretical calculations of neutrino-argon cross sections are not well quantified at these energies, and are expected to be large. In the inelastic NC or CC scattering, the neutrino excites the target nucleus to a low-lying nuclear state, followed by nuclear de-excitation products such as gamma rays or ejected nucleon of a few to tens of MeV that can be detected in the CCM detector. The interaction cross sections for these processes do not scale as $N^2$ and therefore are at least one order of magnitude smaller than that of \cevns process. CCM provides a unique opportunity to measure tens-of-MeV inelastic CC/NC neutrino-argon cross sections that will provide much needed strong constraints on the theoretical models. Such a measurement will enhance DUNE's capabilities of detecting a once-in-a-lifetime large-statistics measurement of supernova neutrinos and will improve its overall supernova neutrino program. 


\bibliography{bibliography}

\end{document}